%% file: ms.tex
\documentclass[12pt,preprint]{aastex}

\usepackage{natbib}

\newcommand{\noprint}[1]{}

\def\cm#1{\ifmmode {\,{\rm cm^{-#1}}}                  
        \else \hbox{$\,${\rm cm$^{\rm -#1}$}}\fi}
\def\raw {\ifmmode\rightarrow\else$\rightarrow$\fi}
\def\ex#1{\ifmmode {\times 10^{#1}}         
        \else \hbox{{$\times 10^{\rm #1}$}}\fi}

\newcommand{\kms}{\mbox{km~s$^{-1}$}}

\newcommand{\mloss}{\mbox{$\dot{M}$}}
\newcommand{\my}{\mbox{$M_{\odot}$~yr$^{-1}$}}
\newcommand{\ls}{\mbox{$L_{\odot}$}}
\newcommand{\msun}{\mbox{$M_{\odot}$}}

\newcommand{\ai}{\mbox{\'{\i}}} 
\newcommand{\vexp}{\mbox{$V_{\rm exp}$}} 
 
\newcommand{\vsys}{\mbox{$V_{\rm sys}$}} 
\newcommand{\vlsr}{\mbox{$V_{\rm LSR}$}}

\newcommand{\teff}{\mbox{$T_{\rm eff}$}}

\newcommand{\hal}{\mbox{H$\alpha$}}
\newcommand{\hb}{\mbox{H$\beta$}}

\slugcomment{To appear in ApJ}

\shorttitle{Gusty post-AGB winds at the core of M\,2-56}
\shortauthors{S\'anchez Contreras et al.}

\begin{document}


\title{Fast, gusty winds blowing from the core of the pre-planetary nebula M\,2-56} 


\author{C. S\'anchez Contreras\altaffilmark{1}, C. Cortijo-Ferrero\altaffilmark{2}, L.F. Miranda\altaffilmark{2,3}, A. Castro-Carrizo\altaffilmark{4}, and 
V. Bujarrabal\altaffilmark{5}}

\altaffiltext{1}{Departamento de Astrof\ai sica, Centro de Astrobiolog\ai a, CSIC-INTA, Ctra de Torrejon 
a Ajalvir, km 4, 28850 Torrejon de Ardoz, Madrid, Spain; csanchez@cab.inta-csic.es}


\altaffiltext{2}{Instituto de Astrof\ai sica de Andaluc\ai a-CSIC, 
C/ Glorieta de la Astronomia s/n E-18008 Granada, Spain; clara@iaa.es, lfm@iaa.es}

\altaffiltext{3}{Departamento de F\ai sica Aplicada, Facultade de Ciencias, 
Universidade de Vigo E-36310 Vigo, Spain}

\altaffiltext{4}{Institut de Radioastronomie~Millim\'etrique, 300 Rue de la Piscine, F-38406 Saint Martin d'H\`eres, France; ccarrizo@iram.fr}

\altaffiltext{5}{Observatorio Astron\'omico Nacional, 
Apartado 112, Alcal\'a de Henares, E-28803 Madrid, Spain;
v.bujarrabal@oan.es}









\begin{abstract}

We report optical long-slit spectra and direct imaging (ground-based
and with $HST$) of the pre-planetary nebula (pPN) M2-56 obtained at
different epochs. The optical nebula is composed by shock-excited
material distributed in two pairs of nested lobes with different
sizes and surface brightness. The compact, bright inner lobes (ILs)
have an angular size of $\sim$1\farcs5$\times$1\arcsec\ each and
display closed, bow-shaped ends.  The extended, faint outer lobes
(OLs), which enclose the inner ones, have an agular size of
$\sim$13\arcsec$\times$10\arcsec. 
Within the ILs and the OLs
the velocity increases with the distance to the center, however, the
ILs show expansion velocities larger than the OLs. 
Consistent with the large speeds reached by the ILs (of up to
$\sim$350\,\kms\ at the tips), we have measured the expansive proper motions of the knots 
($\Delta \theta_{\rm t}$$\sim$0\farcs03 yr$^{-1}$) by comparing $HST$ images taken in
1998 and 2002. Moreover, we have discovered remarkable changes with
time in the continuum and line emission spectrum of M\,2-56. In 1998,
we detected a burst of \hal\ emission from the nebula nucleus that is
interpreted as an indication of a dense, fast ($\sim$350-500\,\kms)
bipolar wind from the nebula's core (referred to as ``F1-wind''). Such
a wind has been recently ejected (after 1989) probably as
short-duration mass-loss event. Our data also reveal an optically
thick compact cocoon (or shell-like structure) and a
\ion{H}{2} region around the central star that result from further
post-AGB mass-loss after the F1-wind. Recent brightening of the
scattered stellar continuum as well as an increase of scattered \hal\ emission along the
lobes is reported, both results pointing to a decrease of the optical
depth of the circumstellar material enshrouding the star.
The data presented here unveil the complex post-AGB mass-loss history
of this object, whose rapid evolution is driven by multiple episodes
of mass outflow, not regularly spaced in time, leading to: ($i$) acceleration of the molecular envelope
that surrounds the optical nebula (kinematical age $t_{\rm
k}$$\sim$1400\,yr -- Castro-Carrizo et al., 2002), ($ii$) the OLs ($t_{\rm k}$$\sim$350-400\,yr), 
($iii$) the ILs ($t_{\rm k}$$\sim$40\,yr), ($iv$) the F1-wind ($t_{\rm
k}$$<$10\,yr), and ($v$) the nuclear cocoon and
\ion{H}{2} region ($t_{\rm k}$$\la$2\,yr?).
The sucessive multiple post-AGB winds in M\,2-56 are characterized by
ejection speeds increasing with time.
In contrast, the mass-loss rate and linear momentum show a time decreasing trend. 

\end{abstract}



 



\section{Introduction} \label{intro}
Planetary nebulae (PNe) evolve from Asymptotic Giant Branch (AGB)
stars after a brief ($\approx$10$^3$\,yr), intermediate stage known as
the pre-Planetary Nebula (pPN) or post-AGB phase. During the AGB to PN
transition dramatic transformations occur: the spherical, slowly
expanding ($V_{\rm exp}$\,$\sim$\,15\,km/s) circumstellar envelope
(CSE) expelled during the AGB becomes a nebula with, usually, clear
departures from sphericity and fast ($\sim$\,100\,km/s) outflows
directed along one or more axes. 
The varied PN and pPN morphologies include
not only axisymmetric (elliptical and bipolar) shells but also
multipolar (multiaxial) structures, multiple co-axial shells, highly
collimated jet-like ejections sometimes arranged in a point-symmetric
structure, etc \citep[e.g.][]{sah98,sahs07,ue07,sio08}. 
Most of the
observed PN and pPN morphologies are difficult to explain without invoking the
action of precessing, perhaps episodic, collimated outflows
\citep[e.g.][]{mir99,gue01,mir06}. In fact, collimated outflows are
considered as the main shaping agent of PNe through their interaction
with the spherical AGB CSE \citep{st98}.

The existence of very young, high velocity bipolar jets in some PPNe
and evolved AGB stars is evidenced, for example, by water maser observations
\citep[e.g.][and references therein]{ima02,ima07,sua09}. In some
pPNs/PNs, fast, post-AGB winds are also revealed by P-Cygni profiles
in the \hal\ (and other recombination lines) emission close to the
central star \citep{san01,san08}.
In spite of the growing evidence of jet-like ejections in PPNs, 
their origin, nature (episodic or continuous?), typical mass-loss rates, life-times, etc, 
are still very poorly known.
Detailed studies of PPNs are crucial to obtain information about the
properties and dynamics of post-AGB winds and their evolution.
Optical spectroscopic observations of pPNs and PNs are particularly
useful for probing the post-AGB winds and their interaction with the
CSE formed in the previous AGB phase.


This work focuses on the pPN M\,2-56 (= IRAS\,23541+7031= PK 118+08),
which so far remains relatively poorly characterized, especially in
the visible. A ground-based optical image obtained by Goodrich (1991,
hereafter G91) showed a $\sim$10\arcsec$\times$2\arcsec\ bipolar
nebula with its long axis oriented in the East-West direction
(P.A.=90\degr). The limited quality of this image (PSF$\sim$1\farcs6
and low S/N) impeded a correct identification of the different nebular
components in M\,2-56 revealed by data presented in this
work. High-angular resolution optical images obtained with the $HST$
have been used as complementary data by
\cite{sio08} and \cite{cc02} although a detailed study based on these
images has not been done so far.  The $HST$ images show that the
bright nebulosity referred to as the ``West lobe'' by G91 is resolved
into two compact lobes ($\sim$1\farcs5$\times$1\arcsec) emerging from
the center.  The emission line optical spectrum of M\,2-56, with
strong forbidden lines of low-excitation ions, indicates major
excitation by shocks
\citep{goo91,tra93,rie06}. The lack of substantial polarization 
in the observed line spectrum implies that the latter is locally
produced in the shock-excited lobes of M\,2-56 and that there is not a
significant contribution by line emission from the central regions
reflected by the nebular dust \citep[''scattered
spectrum'',][]{tra93}.  (In this work we show that this situation has
recently changed). The optical spectrum of M\,2-56 also shows a weak
red continuum that is the light from a B-type central star attenuated
and/or partially reflected by circumstellar dust \citep{coh77}.

A detailed study of the dynamics and morphology of the molecular
envelope of M\,2-56 with high-angular resolution was carried out by
\cite{cc02}. Interferometric CO mapping shows an incomplete
$\sim$30\arcsec$\times$15\arcsec\ hourglass-like envelope surrounding
the optical lobes and similarly oriented to the latter in the plane of
the sky. The inclination of the symmetry axis of the CO envelope with
respect to the plane of the sky is estimated to be
$i$=17\degr$\pm$2\degr, with the East side of the nebula tilted towards
us. The densest molecular material lies at the center, where the
intersection of the two bipolar lobes shapes an equatorial ring/torus
expanding at 7-8\,\kms. Two compact CO clumps are found at the tips of
each lobe expanding axially with velocities of
$\sim$100-200\,\kms. The total mass and linear momentum of the
molecular envelope of M\,2-56 is $M_{\rm mol}$=0.05\,\msun\ and
$P_{\rm mol}$=2\,\msun\kms, adopting a distance to the source of
$d$=2.1\,kpc and a standard value of the CO-to-H$_2$ relative
abundance of X(CO)=2$\times10^{-4}$. The morphology and dynamics of
the bipolar nebula is interpreted as the result of the interaction
between fast, collimated post-AGB ejections and the slow CSE formed
during the AGB phase. The Hubble-type expansion
of the molecular outflow suggests that such an
interaction took place approximately 1400\,yr ago in a relatively
short time, $\la$300\,yr.

In this paper we will adopt a distance to M\,2-56 of $d$=2.1\,kpc
based on the detailed discussion and final estimate by
\cite{cc02}. As estimated by these authors, 
for such a distance, the luminosity of M\,2-56 obtained by integrating
its spectral energy distribution (SED) from the optical to the
mm-wavelength range is $\sim$5500\,\ls.

In this work, we report multi-epoch optical long-slit spectra and
images of M\,2-56 that 
have allowed us to describe the structure,
kinematics, physical conditions, and mass-loss history of this object
with unprecedented detail. 
Our results are discussed in terms of the formation and
evolution of M\,2-56, which is currently in rapid transition from the
AGB to the PN stage and presents clear signs of on-going, variable
wind activity at the nucleus. A summary of our findings
is given in Section \ref{summ}.


\section{Observations and data reduction}
\label{obs}

\subsection{Optical Imaging}
\label{obs_im}

\subsubsection{Ground-based images}

We have obtained broad- and narrow-band images of M2-56 using the Wide
Filed Camera (WFC) at the prime focus of the 2.5m INT of the Roque de
los Muchachos Observatory (La Palma, Spain). Observations were
performed on service time mode on June 13$^{th}$, 2009. The WFC
consists of 4 thinned EEV 2k$\times$4k CCDs.  The CCDs have a pixel
size of 13.5\,\micron\ corresponding to 0\farcs33 per pixel. Our
target was observed with the chip CCD4, which has a field of view of
$\sim$12\arcmin$\times$23\arcmin. Two filters were used: H$\alpha$
($\lambda_{\rm c}$=6568\AA, FWHM=95\AA) and Harris $R$ ($\lambda_{\rm
c}$=6380\AA, FWHM=1520\AA). The integration times were about 530 and
2400 s for the R and H$\alpha$ images, respectively.  The compact,
nuclear regions of M2-56 are saturated in the \hal\ image. The weather
was good during the observations and the seeing was about
1\farcs3-1\farcs5.

Images have been debiased and flat field
corrected following the standard procedure using IRAF\footnote{IRAF is
distributed by the National Optical Astronomy Observatories, which are
operated by the Association of Universities for Research in Astronomy,
Inc., under cooperative agreement with the National Science
Foundation.}. Individual frames have been registered and combined for
each filter to produce final images shown in Fig.\,\ref{f_images}. 

\subsubsection{$HST$ images}
\begin{figure}
\epsscale{0.4}
\plotone{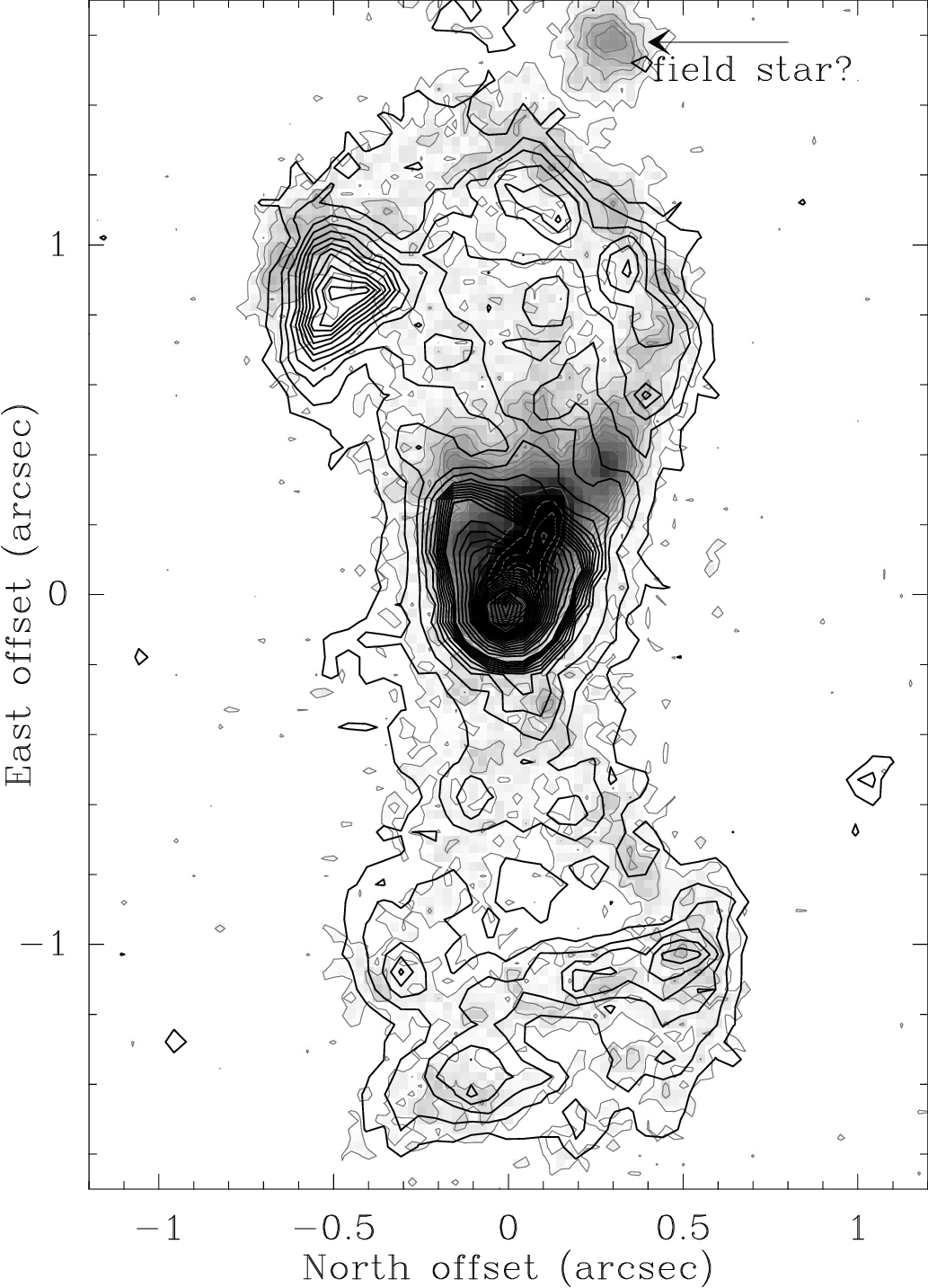}
\caption{$HST$ images of the ILs of M\,2-56 observed in 2002 with the F606W filter
(color scale and blue contours) and in 1998 with the F656N filter
(green contours) showing the proper motions of the tips of the lobes
and other structural differences within their innermost regions
(closest to the center). 
}
\label{f_pm}
\end{figure}

We have used two (\hal\ and continuum) high angular resolution images
of M\,2-56 from the $HST$ archive. The \hal\ image was obtained with
the WFPC2 PC camera using the narrow band filter F656N ($\lambda_{\rm
c}$=6564\AA, FWHM=21.5\AA) on 1998-09-04 (GO 6761; PI:
S. Trammell). This camera has a field of view (FoV) of
40\arcsec$\times$40\arcsec\ and a plate scale of 0\farcs05 per pixel.
Total exposure time was 5000\,s. The continuum ($V$ band) image was
obtained with the ACS/HRC camera using the broad-band filter F606W
($\lambda_{\rm c}$=5907\AA, FWHM=2342\AA) on 2002-12-23 (GO 9463; PI:
R. Sahai). The ACS camera has a field of view of
26\arcsec$\times$29\arcsec\ and a plate scale of 0\farcs025 per pixel.
Total exposure time was 800\,s. The images (Figs.\,\ref{f_images} and
\ref{f_pm}) were pipeline reduced.


\subsection{Optical Spectroscopy}
\label{obs_spec}

We have obtained optical long-slit spectra of M\,2-56 using the
Intermediate Dispersion Spectrograph (IDS) of the 2.5m Isaac Newton
Telescope (INT) and the Andaluc\ai a Faint Object Spectrograph and Camera
(ALFOSC) mounted on the 2.6\,m Nordic Optical Telescope (NOT) of the
Roque de los Muchachos Observatory (La Palma, Spain).  Observations
were performed in three different campaings in 1998, 2000, and
2009. The log of our multi-epoch spectroscopic observations is given
in Table\,\ref{t_log}.

\subsubsection{INT+IDS}
\label{obs_ids}

\begin{figure}
\epsscale{0.75}
\plotone{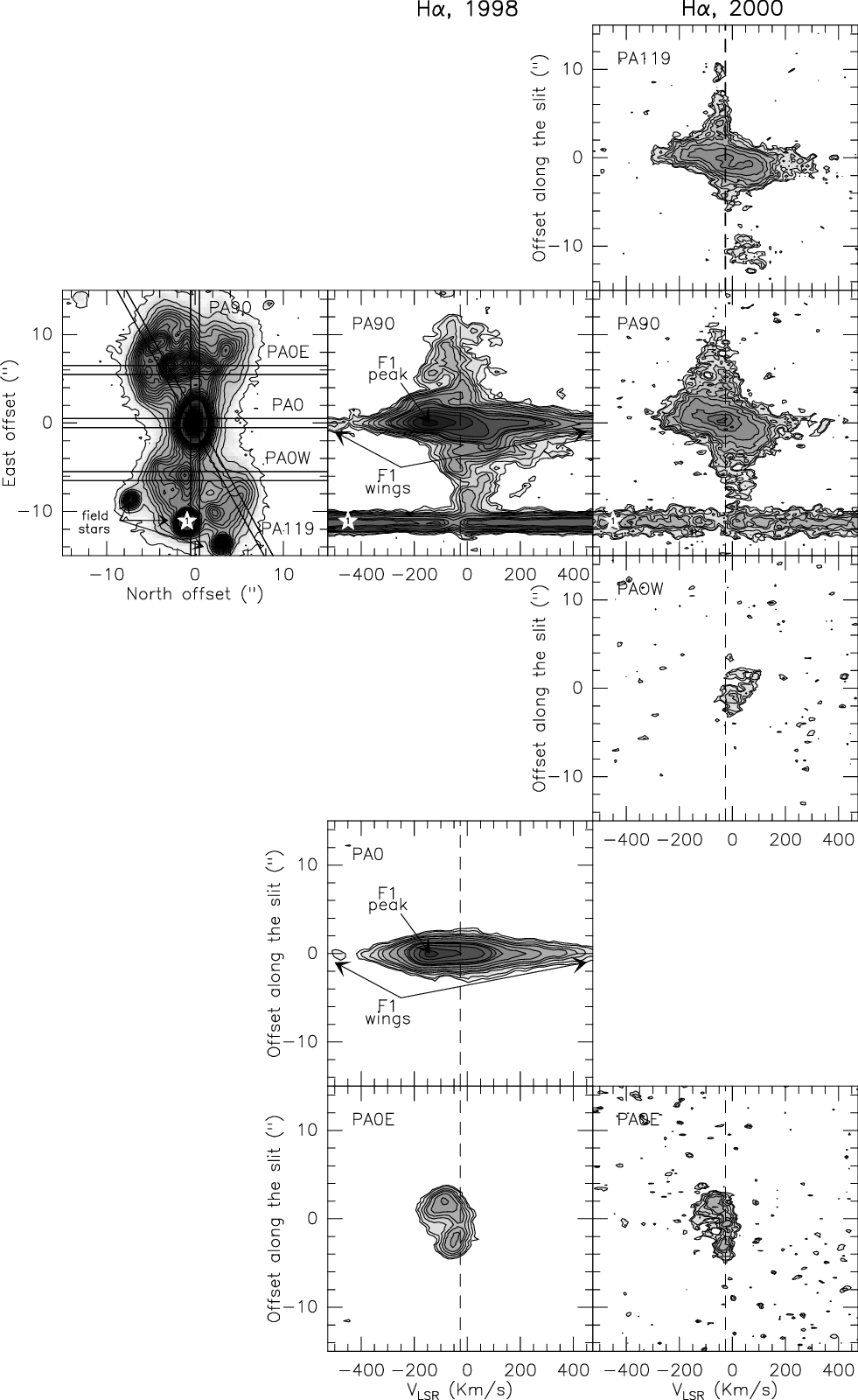}
\caption{Long-slit \hal\ spectra in M\,2-56 for the five slit positions observed in our 1998 and 2000 runs. 
The ground-based \hal\ image of the nebula is shown on the leftmost
panel; slit positions are superimposed on the image.  The origin of
the spatial scale in the spectra coincides with the point of maximum
continuum and \hal\ emission. The LSR systemic velocity of the source,
\vsys=$-$27\,\kms, is indicated by a vertical dashed line on
each spectrum. The intense, blue-shifted emission feature at the
nebula center observed in 1998 and its broad line wings (feature F1)
are indicated by the arrows. We also point the field star that enters
partially slit PA90 and its continuum spectrum (star-like symbol).}
\label{f_spec2}
\end{figure}

For the first set of spectroscopic observations with IDS (run\#1), the
detector used was a TEK 24\,\micron\ CCD with 1124$\times$1124 pixels
mounted on the 500mm camera.
Three slits positions were observed: one at position angle PA=90\degr\
(the symmetry axis of the nebula) passing through the nebula center
(referred to as PA90) and two oriented along PA=0\degr, one passing
through the bright nebula center (PA0) and the other approximately
through the center of the East lobe (at offset $\Delta
\alpha$$\sim$+6\arcsec\ from the center, PA0E) -- see
Fig.\,\ref{f_spec2}, leftmost panel. The slit was 1\arcsec\ wide and long enough to cover the whole nebula.
CuNe lamps were used for wavelength
calibration. The achieved spectral resolution (FWHM of the calibration lamp lines)
was 
$\sim$44\,\kms\ around \hal.

For the second set of spectroscopic observations with IDS (run\#2) the
detector was the EEV10 CCD, with squared pixels of 13.5\micron\
lateral size. Only a clear and unvignetted region of 700$\times$2600
pixels was used (the 2600 pixels were along the spectral axis). The
CCD was mounted on the 500mm camera. A total of four slits positions
were observed with the R1200Y grating (Fig.\,\ref{f_spec2}, leftmost
panel): we repeated positions PA90 and PA0E, already observed in
run\,\#1, and also added two more, one oriented along PA=0\degr\ passing
through the West lobe (at offset $\Delta \alpha$$\sim$$-$6\arcsec\
from the center, PA0W) and another one passing through the nebula
center but oriented along PA=119\degr\ (PA119).
For the slit position PA90 we also observed with the R900V grating. We
used CuNe and CuAr lamps to perform the wavelength calibration. The
spectral resolution achieved was $\sim$\,50\,\kms, at H$\alpha$, and
$\sim$\,83\,\kms, at H$\beta$.  

In both runs, \#1 and \#2, flux calibration was done using sensitivity
functions derived from spectrophotometric standards, namely, HR1544,
HR3454, and HD217086, and taking into account the atmospheric
extinction curve at La Palma.  All observations were performed under
photometric conditions and with a typical seeing of $\sim$1\arcsec\
except for those carried out the first night of our run \#2
(2000-11-11), during which an average seeing of $\sim$2\arcsec\ was
measured. A secondary flux calibration of the mentioned dataset
observed under non-photometric conditions (i.e.\ within the
$\sim$6200-6800\AA\ range covered by grating R1200Y) was performed by
matching the fluxes of the
[\ion{S}{2}]$\lambda\lambda$6716,6731\AA\ doublet lines measured in
1998 and 2000, that is, assuming that the line fluxes have not changed
with time. This is a reasonable assumption given the stability of the
profile of these lines in all our campaigns (including run \#3 in
2009). We derive a value for the scale factor of $\sim$4 
($F^{\rm [SII]}_{2000}$$\times$4$\sim$$F^{\rm [SII]}_{1998}$).
Once applied to all line fluxes in the range 6300-6740\AA, this factor
also corrects for the different seeing in our non-photometric data
($\sim$2\arcsec), which affects the fraction of the nebular emission
that enters the slit.
We double-checked the obtained flux scale factor by comparing the fluxes integrated over
the nebular core (the inner $\sim$2\arcsec) of all lines (not only [\ion{S}{2}]) in the spectra
taken in 2000 along PA90 (under non-photometric conditions) and along
PA119 (under photometric conditions), since both slits pass through
almost the same central regions. 

\begin{figure}
\epsscale{1.3}
\rotatebox{270}{\plotone{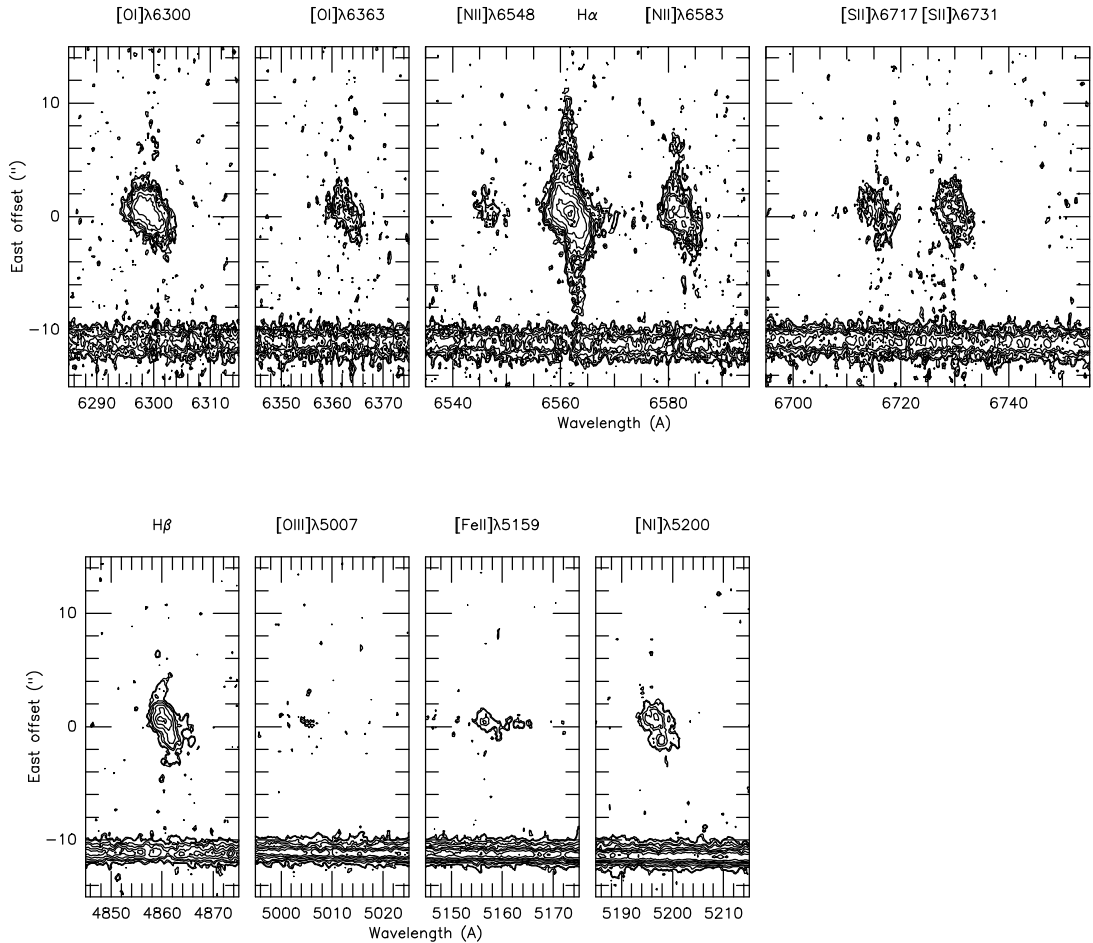}}
\caption{Long-slit spectra over the whole spectral range observed in 1998 and 2000 
for the central slit PA90 (see Table\,\ref{t_log} and Fig.\,\ref{f_spec2}).}
\label{f_spec1}
\end{figure}

The data were reduced using standard procedures for long-slit
spectroscopy within the IRAF package.
The long-slit spectra  presented in Figs.\,\ref{f_spec2} and \ref{f_spec1} have been smoothed
with a flat-topped rectangular kernel of dimension 3$\times$3 pixels
to increase the S/N. 

\subsubsection{NOT+ALFOSC}
\label{obs_alfosc}
\begin{figure}
\epsscale{0.40}
\rotatebox{270}{\plotone{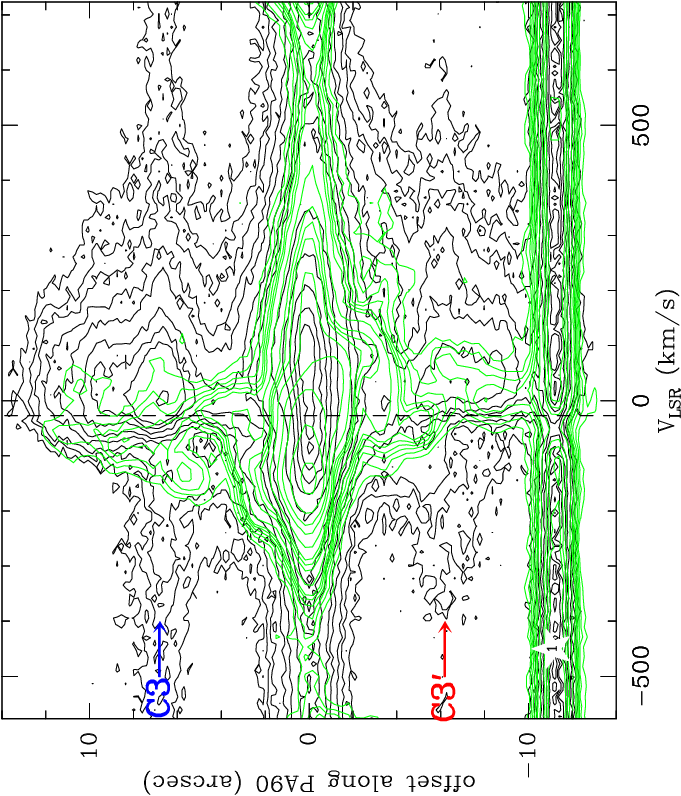}}
\epsscale{0.40}
\rotatebox{270}{\plotone{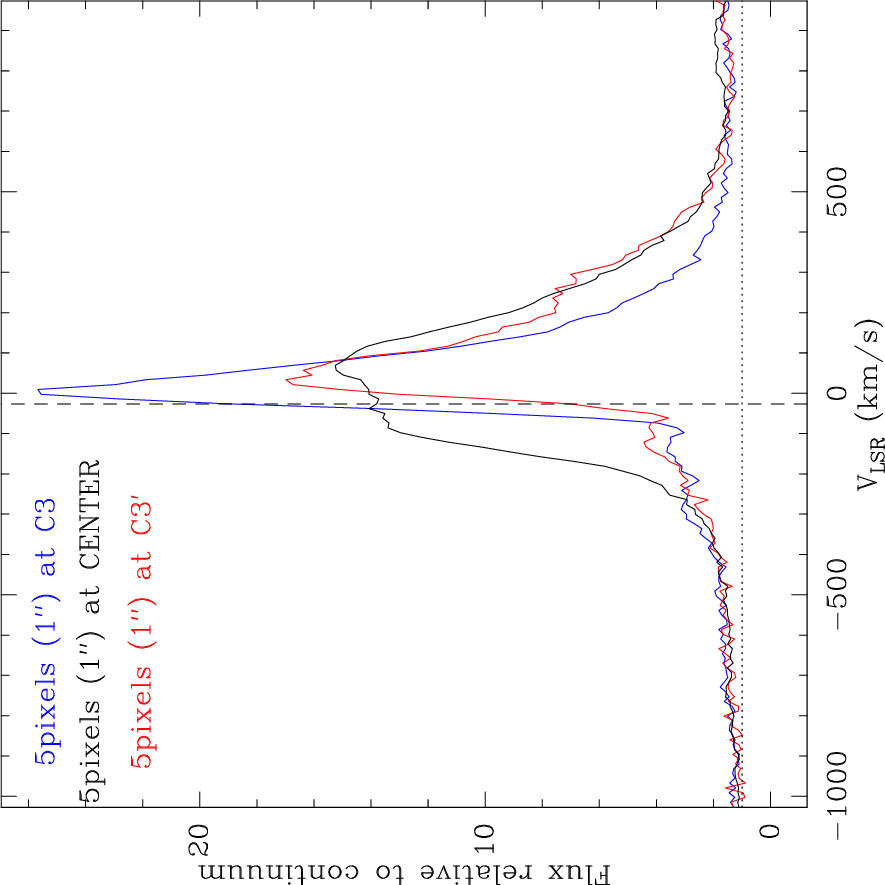}}
\caption{
{\bf Left)} Superposition of the \hal\ long-slit spectrum along PA90
obtained in 2009 (grey color scale) and in 1998 (green contours).  The
systemic velocity is indicated by the vertical dashed line. The loci
of clumps C3 and C3' along the slit is shown.
{\bf Right)} One dimensional \hal\ profile extracted within a
1\arcsec$\times$1\arcsec\ region centered at condensations C3 (blue)
and C3' (red) and at the nebula center (black).  }
\label{f_2009hal}
\end{figure}

\begin{figure}
\epsscale{0.38}
\rotatebox{270}{\plotone{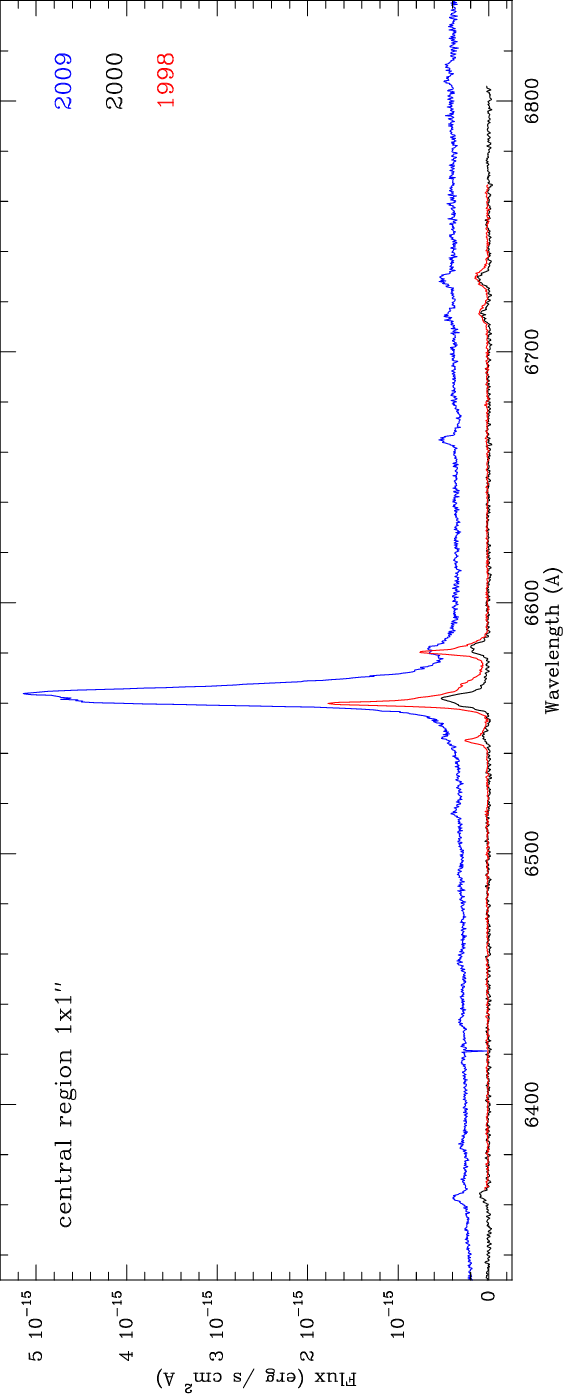}}
\epsscale{0.38}
\rotatebox{270}{\plotone{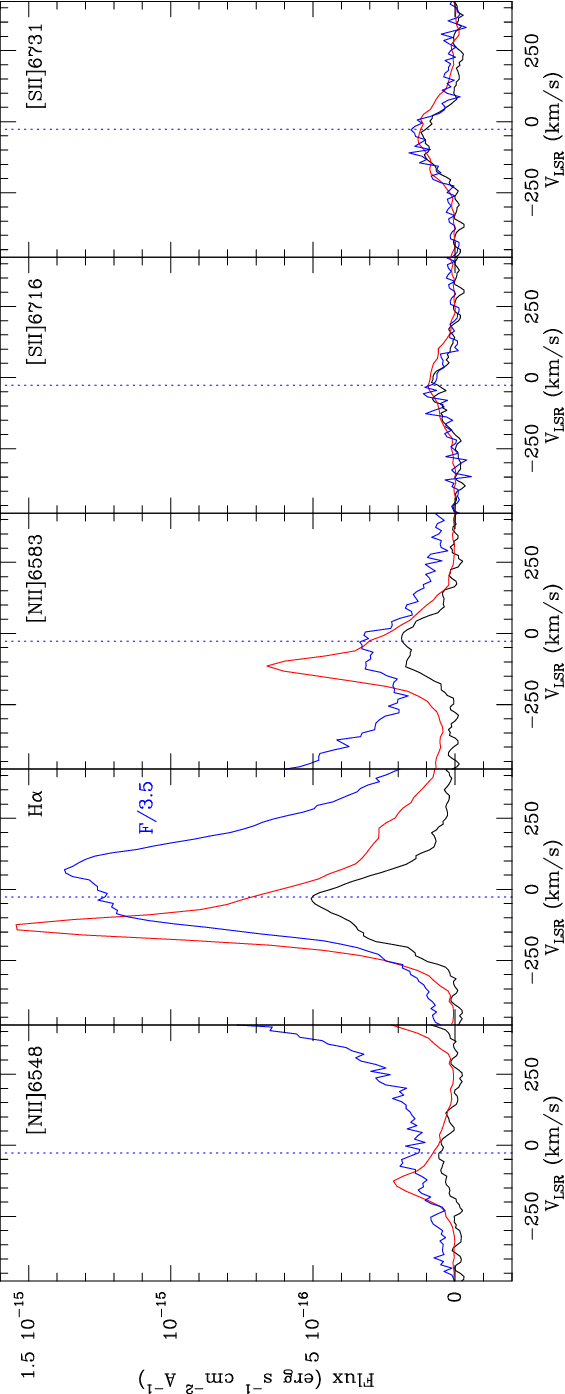}}
\caption{{\bf Top)} Onedimensional spectra from 
a central region of 1\arcsec$\times$1\arcsec\ extracted from our
long-slit spectra PA90 obtained in 1998 (black), 2000 (red), and 2009
(blue). The 1998 and 2000 spectra have been boxcar smoothed to match a
common value of the dispersion of $\sim$0.7\AA. {\bf Bottom)} The same
as in top panel after fitting and subtracting the continuum and for
selected transitions detected in all three epochs using a velocity
scale in the $x$-axis. The \hal\ profile in 2009 has been scaled down
by a factor 3.5 for an easier comparison with the weaker emission
profile in 1998 and 2000.  The systemic velocity is indicated by the
vertical blue-dotted line. Note the similar profile of the
[\ion{S}{2}] doublet lines in all epochs and the remarkable
differences in the \hal\ and [\ion{N}{2}] lines.}
\label{f_1dn}
\end{figure}

Observations were done in service time mode using the CCD\,\#8 detector,
with 2048$\times$2048 13.5\,\micron\ sized pixels, and the VPH Grism
\#17. A 0\farcs9-wide (5\farcm3-long) slit oriented at PA=90\degr\ and
passing through the nebula center (slit PA90) was employed. The
effective spatial resolution (seeing) was 0\farcs9. We observed and
used Ne lamps for wavelength calibration. The achieved spectral
resolution 
$\sim$61\,\kms\ around \hal. The spectrophotometric standard
star BD+284211 was observed and used for primary flux calibration. A
secondary flux scale factor of $\sim$7 has been  
obtained, following the procedure described in the previous subsection
\ref{obs_ids}, and applied to the data. Final spectra, obtained after standard data
reduction with IRAF, are shown in Figs.\,\ref{f_2009all}-\ref{f_1dn}.

\section{Results} 
\label{res}

\subsection{Imaging}
\label{res_im}

\begin{figure}
\epsscale{0.60}
\rotatebox{270}{\plotone{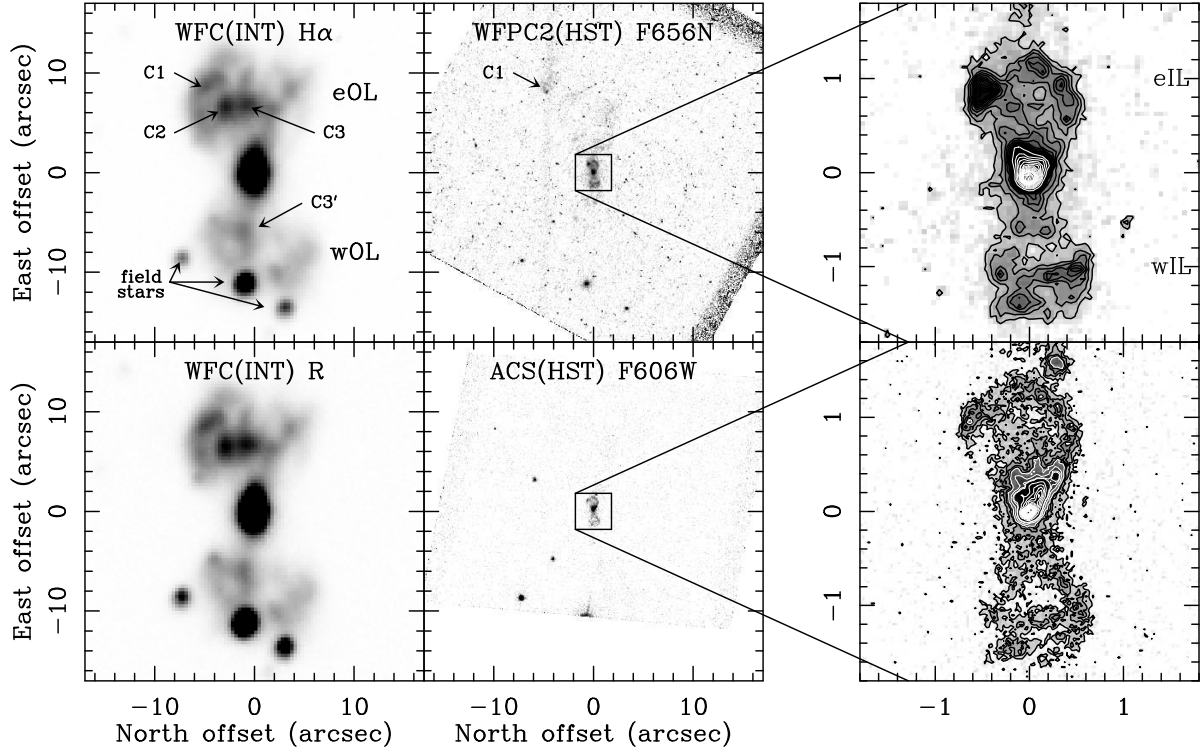}}
\caption{{\bf Left)} Images of M\,2-56 
obtained with the WFC of the 2.5\,m INT through the narrow \hal\ 
and $R$ broad-band filters (top and bottom, respectively) plotted
using a logarithmic scale. The faint, extended east and west outer
lobes (eOL and wOL), nebular condensations C1, C2, C3, and C3', and field stars are labeled. {\bf Middle)} Images obtained with
the $HST$ through the F656N (\hal) and F606W filters (top and bottom,
respectively) plotted using a logarithmic scale. The bright, compact
east and west inner lobes are labeled (eIL and wIL). {\bf Right)}
Inset of the $HST$ images (FoV=3\farcs6$\times$3\farcs6) showing the
small-scale structure of the inner nebular lobes.}
\label{f_images}
\end{figure}

In Fig.\,\ref{f_images} we present images of M\,2-56 that
unveil the different components of the optical nebula and their
morphology. We find two pairs
of nested lobes similarly oriented in the plane of the sky (along
P.A.$\sim$90\degr) but with different sizes and a large contrast in
surface brightness. The bright inner lobes (ILs), which can only be
identified and spatially resolved in the $HST$ images, have an angular
size of $\sim$1\farcs5$\times$1\arcsec\ and display closed, bow-like
ends, which very likely represent shock fronts. A 'three finger'-like
structure at the base of the east IL (within 0\farcs5 from the
core) can be seen in the F606W image.
The faint, more extended ($\sim$13\arcsec$\times$10\arcsec), outer lobes
(OLs), which are probed by our deep ground-based images but are only barely
detected in the $HST$ images, enclose the inner ones. Both
pairs of lobes present an overall axial symmetry, however, they
display a remarkable clumpy structure, including strings of compact
knots and curved features that do not always have an axial
counterpart.
%

The surface brightness of the east lobes is larger than that of
the west ones\footnote{Hereafter, we will refer to the east (west) ILs and OLs as 
eILs and eOLs (wILs and wOLS), respectively.}. This is most likely a result of a larger
extinction of the west side of the nebula, which is receding from us,
given the hourglass distribution of the molecular, and presumably
dust, envelope around the optical nebula (\S\,\ref{intro} and
\S\,\ref{extinction}). The larger brigthness contrast at the central parts
of the nebula (e.g., note that the three-finger structure of the eIL does not have a counterpart in the west) 
is consistent with a larger extinction
along the densest equatorial regions.

In spite of their different filter bandwidth, both the broad- and
narrow-band $HST$ images of M\,2-56 trace approximately the same gas
component, as deduced from our long-slit spectra
(\S\,\ref{res_spec}). First, the contribution of the continuum
to the emission within the F606W and F656N filters was negligible 
at the time the $HST$ images were acquired, except maybe in the
innermost regions of the ILs (within $\pm$0\farcs25) where a very weak
continuum level is detected in our spectra. 
Second,
although the broad-band filter F606W includes several 
lines 
other than \hal, e.g.,
[\ion{N}{1}]$\lambda$5200\AA,
\ion{Na}{1}$\lambda$5893\AA, 
[\ion{O}{1}]$\lambda$$\lambda$6300,6363\AA, 
[\ion{N}{2}]$\lambda$$\lambda$6548,6583\AA\ and
[\ion{S}{2}]$\lambda$$\lambda$6717,6731\AA\ (see also G91), 
the majority of these low-excitation transitions are expected to arise
in the same \hal-emitting regions probed by the F656N filter. The only
exception could be the [\ion{N}{2}] and [\ion{S}{2}] doublets, which may
proceed from regions with a slightly higher excitation, however, these
lines have a contribution to the total line flux within the F606W
filter of less than 30\%\  (see Table\,\ref{t_flux} and G91). 

We have noticed some differences between the two $HST$ images, F656N
and F606W, which were observed in 1998 and 2002, respectively. In
particular, the string of knots/clumps at the tips of the ILs in 2002
is located $\approx$0\farcs10-0\farcs16 ahead the same structure as it
was observed in 1998 (Fig.\,\ref{f_pm}). This is true for both the
east and west ILs and we believe it is due to proper
motions\footnote{Here the term ''proper motion'' is not used to
describe the apparent motion of the nebula as a whole with respect to
more distant stars but to refer to
the tangential motion of material in the expanding nebula (e.g.\ the
knots).} ($\Delta \theta_{\rm t}$$\sim$0\farcs03\,yr$^{-1}$) resulting
from the fast expansion of the ILs (see \S\,\ref{res_9800} and
\ref{inc}).

The base of the eIL also has a quite different appearance in the F606W
and F656N images. In particular, the three finger-like structure
visible in the F606W image is absent in the F656N image. In the latter
image, the same region, i.e.\ the base of the eIL, shows an asymmetrical 
cone-like morphology but has a smoother brightness distribution
slightly elongated to the north-east. These changes cannot be
unambiguously attributed to real variations with time of the nebular
structure since we cannot rule out some contribution from continuum
emission (due to dust scattering in these innermost regions) that may
be different in the F656N and F606W filters. 

Finally, we have compared our two-epoch \hal\ images observed with the
$HST$ in 1998 and with the INT in 2009. There is an emission clump
located at offset $\Delta \alpha$=5\arcsec, $\Delta \delta$=8\arcsec\
that is visible in both data sets (clump C1;
Fig.\,\ref{f_images}). Surprisingly, there is another pair of even
brighter clumps in the ground-based image, labeled C2 and C3, which
are not detected in the $HST$ images.  Other fainter clumps or knots
within the OLs (e.g.\ C3') visible in our 2009 images may have
remained below the noise level in the $HST$ images.

\subsection{Spectroscopy}
\label{res_spec}

In this section, we show the results from our multi-epoch
spectroscopic observations. In Table\,\ref{t_flux} we report the
fluxes ($F_{\lambda}$) of the different lines detected. These fluxes
have been measured from the PA90 spectrum integrating along the
1\arcsec-wide slit and over the full width of the spectral profile,
after fitting and subtracting the continuum level.  In the table, we
also provide dereddened fluxes ($I_{\lambda}$) for a mean logarithmic
extinction coefficient of c(\hb)=1.5$\pm$0.1 (see
\S\,\ref{extinction}). 

We have discovered significant changes with time in the line and
continuum emission spectrum of M\,2-56. Amongst the different lines
detected, \hal\ is the transition displaying the most extreme profile
variations. We will first report our results obtained from the spectra
observed in 1998 and 2000, for which the differences found only affect
the line emission from the nuclear regions. Next, we will describe the
long-slit spectrum along PA90 obtained in 2009, which shows additional
changes with respect to the data obtained in our two previous runs.

\subsubsection{1998 and 2000 observations}
\label{res_9800}
In Figs.\,\ref{f_spec2} and \ref{f_spec1}, we show the optical
spectrum of M\,2-56, which is composed of recombination and forbidden
line emission plus a weak, red continuum. In 1998, the continuum
emission arises from a central compact region that is unresolved in
our data (with a PSF$\sim$0\farcs9).
The continuum intensity near \hal\ (around 6510\AA) in 1998 integrated
over the 1\arcsec$\times$1\arcsec\ central region of M\,2-56 is
$\sim$1.5($\pm$0.4)$\times$10$^{-17}$\,erg\,s$^{-1}$\,cm$^{-2}$\,\AA$^{-1}$.
In 2000, 
we derive an upper limit for the continuum intensity within the same central region
of $<$2.4$\times$10$^{-17}$\,erg\,s$^{-1}$\,cm$^{-2}$\,\AA$^{-1}$
(i.e.\, $<$2\,$\sigma$). Therefore, the non detection of the continuum
in 2000 is consistent with its intensity remaining constant from 1998
to 2000.
(As shown in \S\,\ref{res_09}, the continuum is brighter in 2009). 

The long-slit \hal\ spectrum is characterized by a wide profile with a
quite complex spectral and spatial distribution
(Fig.\,\ref{f_spec2}). Along the nebula axis (slit PA90), the \hal\
emission can be traced from the brightest central regions up to the
fainter tips of the OLs at $\pm$12\arcsec.
The \hal\ emission from the east and west lobes (approaching to and
receding from us, respectively,
\S\,\ref{intro}) is, on average, blue- and red-shifted, consistent
with an expansive nebular kinematics. The spatial distribution of the
emission from the bright, compact ILs is only partially resolved in
our long-slit spectra and can be traced from the center up to
$\pm$2\arcsec. In these innermost regions, the much fainter OLs are
not expected to contribute significantly to the \hal\ emission given
the large brightness contrast with the ILs evidenced by the $HST$
images.  Along the ILs, the line-of-sight or radial expansion velocity
increases with the distance reaching a maximum value of $V_{\rm
r}$=\vlsr$-$\vsys$\sim$110\,\kms\ at the tips. In these outermost
regions of the ILs, the \hal\ profile is rather broad, with a full
width at zero intensity (FWZI) of $\sim$360\,\kms. The
\hal\ long-slit profile along and across the OLs (e.g., along PA0E)
is roughly consistent with the lobes being expanding bubble-like
structures. We measure a value for the mean radial expansion velocity
of the OLs of $V_{\rm r}$$\sim$50\,\kms, which is smaller than that
derived for the ILs. In the outermost parts of the OLs, at axial
offsets $\pm$10\arcsec-12\arcsec\ (PA90), our spectra indicate a
slightly smaller radial expansion velocity of $V_{\rm r}$=30-35\,\kms.

We have found remarkable differences between the \hal\ emission from
the nucleus of M\,2-56 as observed in 1998 and in 2000
(Fig.\,\ref{f_spec2}). In 1998: a) the \hal\ profile shows emission
wings that are much broader (FWZI$\sim$1000\,\kms) than in 2000
(FWZI$\sim$450-500\,\kms) -- this difference is not explained by the
lower S/N achieved in 2000 (note that the fainter emission from the
OLs is well probed in our 2000 spectrum); and b) the \hal\ emission
peak is blue-shifted by $\sim$110\,\kms\ from the systemic velocity,
unlike the spectrum in 2000, which peaks at \vsys. {\sl We refer to
the intense, blue-shifted emission feature and its broad wings
observed in 1998 as feature F1}. We have checked that feature F1 was
not present either in the long-slit spectrum along the nebula axis
acquired in 1989 by G91.  The broad wings of feature F1 are not
symmetric, the blue wing seems to be depressed relative to the red
one, which may be interpreted as a blue-shifted (P-Cygni like)
absorption in the profile. The blue-shift of the wing absorption is
found at $V_{\rm r}$$\sim$$-$430\kms. The presence of the spectral
component F1 is clearly appreciated also in the \hal\ spectrum taken
with the slit oriented along the nebula equator PA0. The angular size
of the region that produces feature F1 is very small, $\sim$0\farcs5
along the main symmetry axis of the nebula and unresolved along the
equator ($<$0\farcs6).

In 1998, feature F1 was not only visible in the \hal\ spectrum but
also in the [\ion{N}{2}]$\lambda\lambda$6548,6584\AA\ lines. In
contrast, feature F1 is not present in the
[\ion{S}{2}]$\lambda\lambda$6716,6732\AA\ transitions: note that the
two lines of the doublet have comparable profiles in
1998 and 2000 (see Fig.\,\ref{f_1dn}). The
profiles of the [\ion{S}{2}] lines are indeed very much alike to those
of \hal\ and the rest of the transitions as observed in 2000
(and in 2009, see next subsection \S\,\ref{res_09}).

\subsubsection{2009 observations}
\label{res_09}

\begin{figure}
\epsscale{0.31}
\rotatebox{270}{\plotone{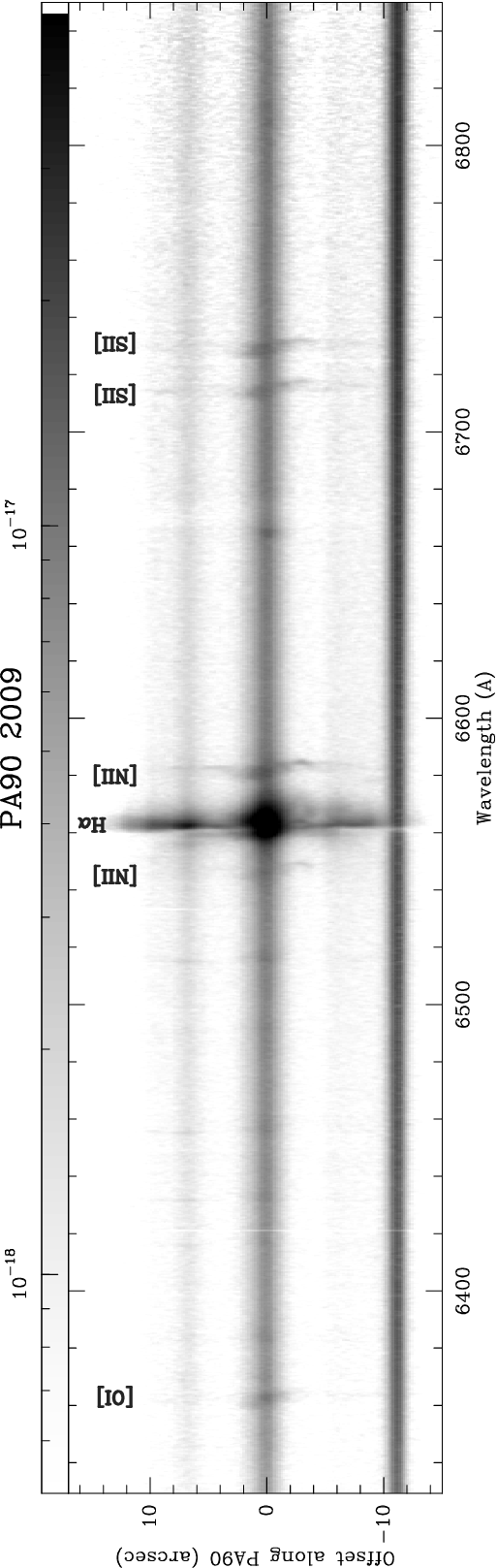}}
\epsscale{0.27}
\rotatebox{270}{\plotone{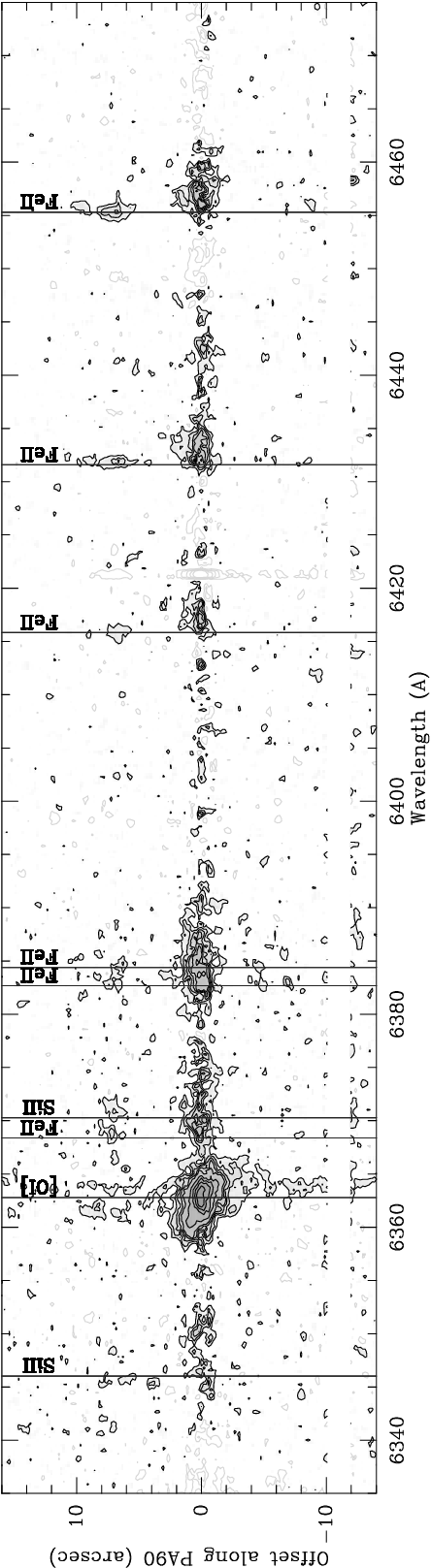}}
\rotatebox{270}{\plotone{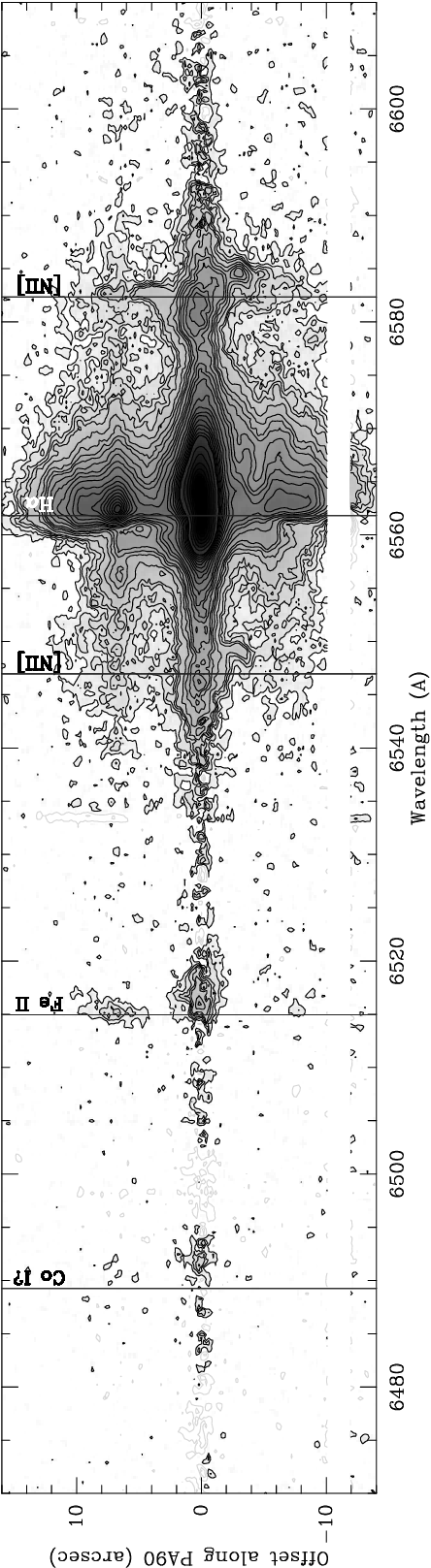}}
\rotatebox{270}{\plotone{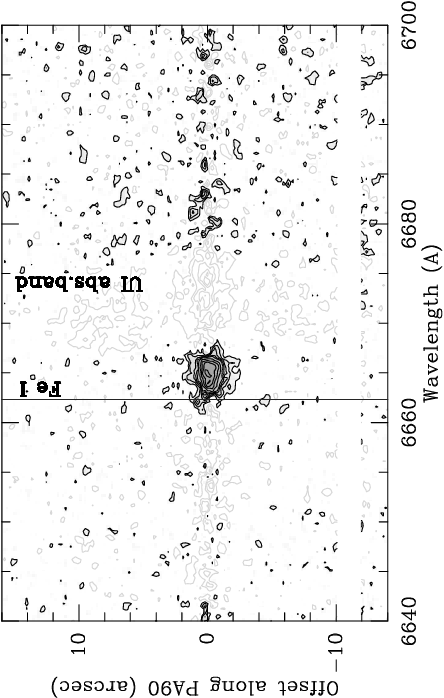}}
\hspace{0.5cm} \rotatebox{270}{\plotone{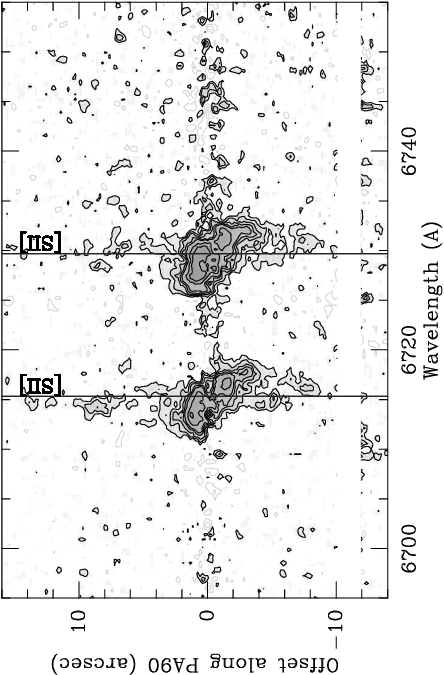}}
\caption{{\bf Top panel)} Long-slit spectrum for slit PA90 obtained in 2009 in a logarithmic color scale. 
The spectrum has been smoothed with a flat-topped rectangular kernel
of dimension 3$\times$3 pixels. The units of the wedge at the top of
the box are erg\,s$^{-1}$\,cm$^{-2}$\,\AA$^{-1}$\,arcsec$^{-1}$.  {\bf
Middle and bottom panels)} Same as above but continuum substracted and
showing selected wavelength ranges. Lines detected are labeled (see
Table\,\ref{t_flux}). The blue vertical lines are placed at \vsys\ for
each of the observed transitions. Contour levels are: 2$\sigma$,
4$\sigma$, 6$\sigma$, 8$\sigma$, 10$\sigma$, 15$\sigma$, 20$\sigma$,
25$\sigma$, 30$\sigma$, 40$\sigma$, 50$\sigma$, 60$\sigma$, from
75$\sigma$ to 300$\sigma$ in steps of 25$\sigma$, from 300$\sigma$ to
1000$\sigma$ in steps of 100$\sigma$ (black), and
$-$3$\sigma$,$-$6$\sigma$,$-$9$\sigma$... (magenta) with
$\sigma$=1.2$\times$10$^{-19}$
erg\,s$^{-1}$\,cm$^{-2}$\,\AA$^{-1}$\,arcsec$^{-1}$ in all panels
except for the last one (containing the [SII] doublet) for which
$\sigma$=1.7$\times$10$^{-19}$
erg\,s$^{-1}$\,cm$^{-2}$\,\AA$^{-1}$\,arcsec$^{-1}$. A small region
(1\farcs7) around the field star at offset $\sim$$-$11\arcsec\ has
been masked for clarity.}
\label{f_2009all}
\end{figure}

Figs.\,\ref{f_2009all} and \ref{f_2009hal} display the long-slit spectrum of M\,2-56
obtained in 2009. Multi-epoch spectra extracted over the central
1\arcsec$\times$1\arcsec\ region are compared in Fig.\,\ref{f_1dn}.

In 2009, a relatively faint, red continuum is observed not only at the
center (as in 1998) but also along the lobes of M\,2-56. As for most
pPNs, the continuum emission is dominated by the stellar photospheric
continuum scattered off by the nebular dust (we have checked that the
contribution by nebular continuum is negligible given the intensity of
the \hal\ emission).
The spatial distribution of the scattered continuum peaks at the
nebula center and has two relative maxima at axial offsets
$\pm$6\farcs5. The latter regions correspond to condensations C3 and
C3' in the ground-based images (Fig.\,\ref{f_images}). The central
region where the continuum is observed has a deconvolved angular size
along the axis of $\sim$1\farcs5, i.e.\ significantly larger than in
1998 ($<$0\farcs5).
The continuum flux near \hal\ (around 6510\AA) within the
1\arcsec$\times$1\arcsec\ central region in 2009 is
$\sim$3.2$\times$10$^{-16}$\,erg\,s$^{-1}$\,cm$^{-2}$\,\AA$^{-1}$,
that is, larger than that measured in 1998 by a factor $\sim$20.  

The shape of the \hal\ long-slit profile shows marked differences
with respect to that observed earlier, not only in the central regions
but also along the extended OLs. The \hal\ emission from both the
approaching and receding (west and east, respectively) lobes is on
average red-shifted, in contrast to what we observed in 1998 and
2000. The red-shift of the \hal\ emission from the aproaching eOL is
smaller than that from the receding wOL.
At the nebula center, the \hal\ emission has a broad asymmetric
profile that peaks at \vlsr=65\kms, i.e. red-shifted (by $\sim$90\kms)
with respect to \vsys. The FWZI of the \hal\ nuclear profile cannot be
accurately determined because of blending with the adjacent
[\ion{N}{2}] doublet; we estimate that it is larger than 1300\kms\ and
probably reaches up to 2500\kms.

The observed red-shift of the \hal\ line for both the approaching and
receding lobes indicates that in 2009 a significant fraction of the
\hal\ emission in the lobes is scattered, i.e.\ is not locally produced in the lobes 
but rather arises at the nebular core and is reflected by the nebular
dust. In fact, the similar spatial distribution of the \hal\ emission
and the reflected continuum beyond $\pm$6\arcsec\ indicates that in
these regions the \hal\ emission is mainly scattered. 
In the inner regions,
however, some contribution of unscattered \hal\ emission (produced
locally in the lobes) can still be appreciated. 
Composite \hal\ line profiles consisting of two components (scattered and unscattered) 
are also found in other pPNs, for example, 
M\,1-92 \citep{tra93}, M\,2-9 \citep{solf00}, CRL\,618 \citep{tra93,san02}, etc.  

In contrast to \hal, the long-slit profiles of the forbidden
transitions [\ion{O}{1}]$\lambda$6363\AA, the
[\ion{N}{2}]$\lambda\lambda$6545,6584\AA\ and
[\ion{S}{2}]$\lambda\lambda$6716,6731\AA\ in 2009 are totally consistent with
those observed in our previous runs\footnote{Except for the presence
of feature F1 in the [\ion{N}{2}] doublet lines in 1998, \S\,\ref{res_9800}}: roughly
point-symmetric with respect to the spatial origin and \vsys\ and with the
emission from the east and west lobes blue- and red-shifted,
respectively. This is
consistent with the forbidden lines being predominantly produced
locally in the shocked nebular material. 

We have discovered a number of emission lines in this work, including 
several \ion{Fe}{2}, \ion{Fe}{1}, and \ion{Si}{2} permitted transitions
(Fig.\,\ref{f_2009all} and Table\,\ref{t_flux}). 
We have also detected a broad (FWHM$\sim$6\AA) absorption feature
centered near 6673\AA\ at the nebula center but
slightly blue-shifted, near 6670\AA, along the east and west
lobes. The relative blue-shift of the band between the center and the
lobes indicates that it does not have an interstellar or telluric
origin but rather is associated to M\,2-56. We have not found a
satisfactory identification for this broad absorption, which may be a
solid state or ice feature given its large width; a blend of atomic
lines cannot be ruled out either.

The emission distribution of the recombination lines discovered is
quite different from that of the forbidden transitions. As for \hal,
the former are red-shifted with respect to \vsys\ along the approaching eOL as
well as at the nebula center (these transitions are not detected in
the fainter west lobes). The red-shift measured in the eOL is smaller
than that in the center. We conclude that the observed recombination
lines most likely arise at the nebula nucleus and their emission is
scattered by circumstellar dust.


Another important change detected in 2009 
is the increase of the relative intensity of \hal\ with respect to the
various forbidden transitions observed.
For example, in 2009,
the \hal-to-[\ion{N}{2}]$\lambda$6584\AA\ intensity ratio reaches a
value of $\sim$15, much larger than in earlier epochs,
$\sim$3.2-3.4. 
The large \hal-to-[\ion{N}{2}]$\lambda$6584\AA\ intensity ratio is
consistent with the presence of a compact \ion{H}{2} region at the
heart of M\,2-56 responsible for the extra \hal\ flux in 2009: note
the brightnening of the \hal\ line by a factor $\approx$10 relative to
the flux measured in 2000.

\section{Nebular components and their spatio-kinematic structure}
\label{components}

In this section we describe the main nebular components in M\,2-56
revealed by our data and analyze their spatio-kinematical
structure. We also estimate the inclination of the nebula axis, which is
a critical parameter to correct for projection effects both in the
geometry and expansion velocities observed.

\subsection{The inclination of the nebula}
\label{inc}

The inclination of the nebula axis is derived in two different
ways. First, from the proper motions observed in the $HST$ images
(Fig.\,\ref{f_pm} and \S\,\ref{res_im}). For a distance to M\,2-56 of
$d$=2.1\,kpc (\S\,\ref{intro}), the observed proper motions ($\Delta \theta_{\rm
t}$$\sim$0\farcs03\,yr$^{-1}$) imply a tangential expansion velocity (i.e.\
projected on the plane of the sky) of $V_{\rm
t}$$\sim$325\,\kms. Since the line-of-sight or radial velocity at the
tips of the lobes measured from our long-slit spectra
is $V_{\rm r}$$\sim$110\,\kms\ (\S\,\ref{res_spec}), we obtain an
inclination angle to the plane of the sky of $i$=tan$^{-1}$($V_{\rm
r}$/ $V_{\rm t}$)=19\degr. This value is in very good agreement with
the independent estimate of $i$=17\degr\ for the symmetry axis of the
bipolar CO envelope by
\cite{cc02}.  
We derive an absolute (deprojected) expansion velocity at the tips of
the ILs of $\sim$350\,\kms. This value is similar to the FWZI of \hal\
emission at the ends of the ILs; the latter is known to be a 
reliable indicator of the shock velocity for bow-type shocks, independent of orientation angle, preshock density, 
bow shock shape, and preshock ionization stage \citep{har87}. 

An independent estimate of $i$ can be obtained from the radial
velocities measured at clumps C3 and C3' in the east and west OLs,
respectively, where the \hal\ emission 
is mainly scattered (\S\,\ref{res_09}). 
Assuming that regions C3, C3', and the nebula center are aligned
along a given axis, which is inclined to the plane of the sky by $i$,
and that the clumps move away from the central star at the same
expansion velocity, \vexp, then
we can readily find \vexp\ and $i$ from the following expressions: 

\begin{eqnarray} 
V_{\rm exp}= \frac{V_{\rm r}(C3)+V_{\rm r}(C3')}{2} \\
\sin (i)=\frac{[V_{\rm r}(C3')-V_{\rm r}(C3)]}{[V_{\rm r}(C3')+V_{\rm r}(C3)]}
\end{eqnarray} \\

\noindent
\citep[see, for example,][]{solf00}. Taking into account the values of $V_{\rm r}$(C3)=31.3\,\kms\ and
$V_{\rm r}$(C3')=60.9\,\kms\ measured on the \hal\ spectrum along PA90
(Fig.\,\ref{f_2009hal}), we obtain \vexp=46\,\kms\ and $i$=18.8\degr.
The value of $i$ found is in excellent agreement with that 
computed for the ILs from the proper motions and that of the bipolar
molecular envelope.

\subsection{Spatio-kinematic model of OLs and ILs}
\label{model}

In order to investigate the structure and velocity field of the
optical lobes of M\,2-56 we have developed a simple spatio-kinematical
model that reproduces the 2D profile of our long-slit
spectra. Although the model has been mainly applied to the PA90
spectrum, we have checked that it also qualitatively explains the
spectra at the other slit orientations. Our model calculates the
position-velocity (p-v) diagram assuming a certain morphology for the
envelope (compatible with our direct images) and a given velocity
field, and also taking into account the inclination of the nebula to
the plane of the sky ($i$=18\degr). We note that our
code is not an emission model, that is, does not
predict the \hal\ brightness distribution along PA90. 

\begin{figure}
\epsscale{0.35}
\rotatebox{270}{\plotone{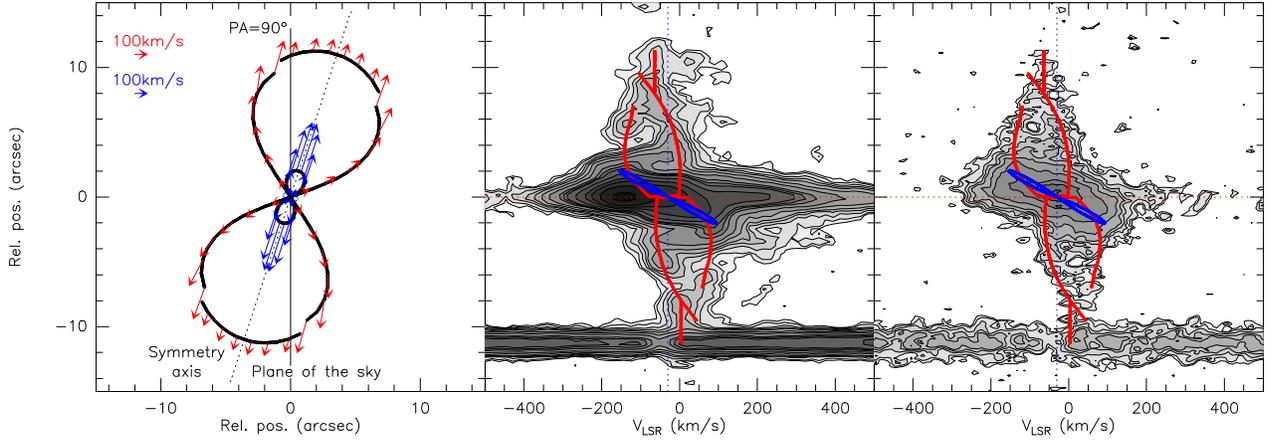}}
\caption{{\bf Left)} Schematic geometry of M\,2-56 
deduced from the model fitting (\S\,\ref{model}). This plot represents
a cut of the nebula by a plane perpendicular to the plane of the sky
in the direction of PA90. In this diagram 1\arcsec\ corresponds to
3.14$\times$10$^{16}$\,cm ($d$=2.1\,kpc, \S\,\ref{intro}).  The
velocity field in the ILs (OLs) is indicated by blue (red) arrows,
respectively. {\bf Right)} \hal\ spectrum along PA90 slit orientation as observed in 1998 and
2000. The position-velocity diagrams resulting from the model fitting
are superimposed. Contours and color scale are the same as in
Fig.\,\ref{f_spec2}.}
\label{f_modelo}
\end{figure}

In Fig.\,\ref{f_modelo} we present the geometry adopted and synthetic
p-v diagrams predicted by our best model for the \hal\ line. The
nebular component responsible for the spectral feature F1 is not
included in our model because it is not spatially resolved in our
spectra.
We are assuming in our model that the lobes are hollow and
show an overall axial symmetry, as suggested by the optical appearance
of M\,2-56.
For the OLs and the ILs, the model fit yields a linear size of about
(4$\times$10$^{17}$\,cm)$\times$(3$\times$10$^{17}$\,cm) and
(6$\times$10$^{16}$\,cm)$\times$(4$\times$10$^{16}$\,cm),
respectively, measured along the symmetry axis and the perpendicular
direction where the lobes reach their maximum extension.

For the adopted geometry, the velocity field that best reproduces
the long-slit spectrum of the OLs consists of three main components:
1) radial expansion with the modulus of the velocity at each point
increasing linearly with the distance to the nebula center; 2) axial
expansion with the velocity modulus increasing linearly with the
distance to the equatorial plane; and 3) shear-flow kinematics, with
the velocity vectors tangential to the curved surface adopted for the
lobes and with a constant velocity modulus.  The components that
dominate the velocity field in M\,2-56 vary from the base of the OLs,
where components 1) and 3) prevail, to the external regions, where the
velocity field is essentially axial.  
The velocity
field described above, with the flow speed increasing with the
distance to the center, works fine for most parts in the OLs, however,
is not able to reproduce the relatively low radial velocities measured
at the tips, $\sim$110\,\kms\ around $\pm$10-12\arcsec, which implies
an abrupt slow down of the flow. In these outer regions, represented
in our model as bow-shaped caps detached from the main body of the
OLs, the expansion seems to be dominated by a constant velocity axial
field.

Since the spatial structure of the ILs is only partially resolved in
our 2D spectra, we have adopted a simple axial velocity field that is
able to reproduce the overall velocity gradient observed within the
inner $\pm$2\arcsec. We deduce a rather steep variation of
the deprojected expansion speed, increasing from tens of \kms\ near
the center to $\sim$350\,\kms\ at the lobe tips. 

\subsection{The wind responsible for the transient \hal\ feature F1}
\label{f1wind}

We believe that the \hal\ emission burst observed in 1998, referred to
as feature F1, is the consecuence of a fast, dense, and short-lived
post-AGB wind recently ejected (\S\,\ref{dis}). 
We have measured a total angular size along the nebula axis of the
region emitting feature F1 of $\sim$0\farcs5 (1.57$\times$10$^{16}$\,cm).
The spectral profile of F1, notably asymmetrical, peaks at
V$_{\rm r}$\,$\sim$110\,\kms, i.e. bluewards of \vsys. This is consistent
with a compact bipolar flow oriented similarly to
the ILs and OLs,  
the red-shifted emission from the receding (west) side of the F1-wind
being significantly extinguished by the optically thick equatorial
regions of M\,2-56.  A larger extinction towards the inner regions of
the wIL (compared to the base of the eIL) is, in fact,
inferred from the brightness contrast between these two regions
observed in the $HST$ images and is also expected from the
equatorially enhanced density distribution of the molecular envelope
that surrounds the optical nebula of M\,2-56
(\S\,\ref{intro} and \S\,\ref{extinction}). (A one-sided collimated F1-wind is not
impossible but improbable given the bipolar morphology, approximately
symmetric with respect to to the equator, of the optical and CO envelopes of
M\,2-56). Adopting a bipolar flow with an inclination to the plane of
the sky of $i$=18\degr, we obtain a linear extension for this
component of $l_{\rm F1}$$\sim$1.65$\times$10$^{16}$\,cm. The
blue-shift of the intensity peak of F1 indicates that the bulk of the
material in the F1-wind expands at a mean velocity of V$_{\rm
exp}$$\sim$110/sin(18\degr)$\sim$360\,\kms, very similar to the expansion speed at the tips
of the ILs. The broad wings of feature F1 and the blue-shift (P-cygni like)
absorption
of the nuclear \hal\ profile indicate smaller amounts of
gas expanding at larger velocities (up to $\sim$500\,\kms) in the F1-wind. 


\subsection{The central \ion{H}{2} region}
\label{res_hii}

The presence of a compact \ion{H}{2} region around the central star of M\,2-56 is inferred from our 
data (\S\,\ref{res_09}). 
In 2009, a fraction of the \hal\ emission from such a region escapes
the stellar vecinity and reaches the ILs and OLs where it is scattered
off by dust. The kinematics of the \ion{H}{2} region cannot be
straightforwardly determined from the \hal\ line profile since the
emission from this region is seen only after being reflected by the
nebular dust, the resulting line profile being significantly affected
by the distribution and kinematics of the dust.
The broad profile of the \hal\ emission from the nebula nucleus
(FWZI$>$1900-2500\kms) may indicate rapidly outflowing material in the
stellar neighborhood, with very large expansion velocities of up to
$\approx$1200\,\kms, however, we cannot rule out other line broadening
mechanisms (e.g., Ramman scattering) being totally or partially
responsible for the large FWZI of the \hal\ transition \citep[see
e.g.][and references therein]{san08}.

The \hal\ scattered profile observed at C3' and C3 represents the
profile emitted by the \ion{H}{2} region as seen from the dust clumps
located at those positions, i.e.\ from a ''pole-on'' view.
As we can see in Fig.\,\ref{f_2009hal}, the \hal\ 
profiles at C3 and C3' are very similar (except for an overall
relative red-shift of 30-40\,\kms): they have a narrow intense core plus 
broad wings and show a notable P-cygni like profile, with the blue
side depressed. The \hal\ emission towards the center exposes the
profile emitted by the \ion{H}{2} region as seen 
along the
line-of-sight, i.e.\ from a direction nearly perpendicular to the lobes. 
Its overall red-shift indicates a significant contribution
of scattered \hal\ emission also towards the nebula center. In this
direction, the \hal\ profile is broader than that observed towards
clumps C3 and C3' and does not exhibit prominent blue-shifted
absortion. 
This suggests that the P-cygni like absorption observed
at C3 and C3' is produced by neutral or partially neutral material 
outside the \ion{H}{2} region and very likely 
distributed in a bipolar flow.

\section{Extinction}
\label{extinction}

We have used the \hal\ to \hb\ flux ratio measured in 2000
(Table\,\ref{t_flux}) to estimate the extinction in M\,2-56. Since the
\hal\ and \hb\ fluxes have been obtained integrating spatially the spectrum along
slit PA90 over the whole nebula, the value derived represents an average of the extinction along
the lobes, which is expected to decrease from the center (nebular
equator) to the tips of the lobes.
We have measured a Balmer decrement of
\hal/\hb$\sim$12$\pm$1, for which we obtain a logarithmic extinction
coefficient $c(\hb)$=1.5$\pm$0.1 adopting in intrinsic ratio
\hal/\hb$\sim$3.8. The latter value of the intrinsic \hal/\hb\ ratio is expected for the 
shock-excitation conditions in M\,2-56 derived by several
authors comparing observational line relative intensities with the
predictions of theoretical shock excitation models \citep[][and
references therein]{goo91,tra93,rie06}. (For an intrinsic Balmer
decrement closer to the recombination value, \hal/\hb$\sim$3, we
obtain c(\hb)=1.7-2.0.)
From the value of $c(\hb)$, we deduce an average optical
depth near \hal\ (around 6563\AA) of $\tau_{\rm 6563}$$\sim$2.3 and,
therefore, an average extinction in the $V$ band of A$_{\rm
V}$$\sim$3\,mag, 
using the extinction law parametrization by \cite{car89} and assuming
the ratio of total to selective absorption, $R_{V}$, equal to 3.1. Our
result is in good agreement with previous estimates of A$_{\rm V}$=3.3
and 4.8\,mag \citep{goo91,coh77}, especially taking into account the
lower intrinsic Balmer decrement, 3 and 2.85, adopted in these earlier
works.

\begin{figure}
\epsscale{1.}
\plottwo{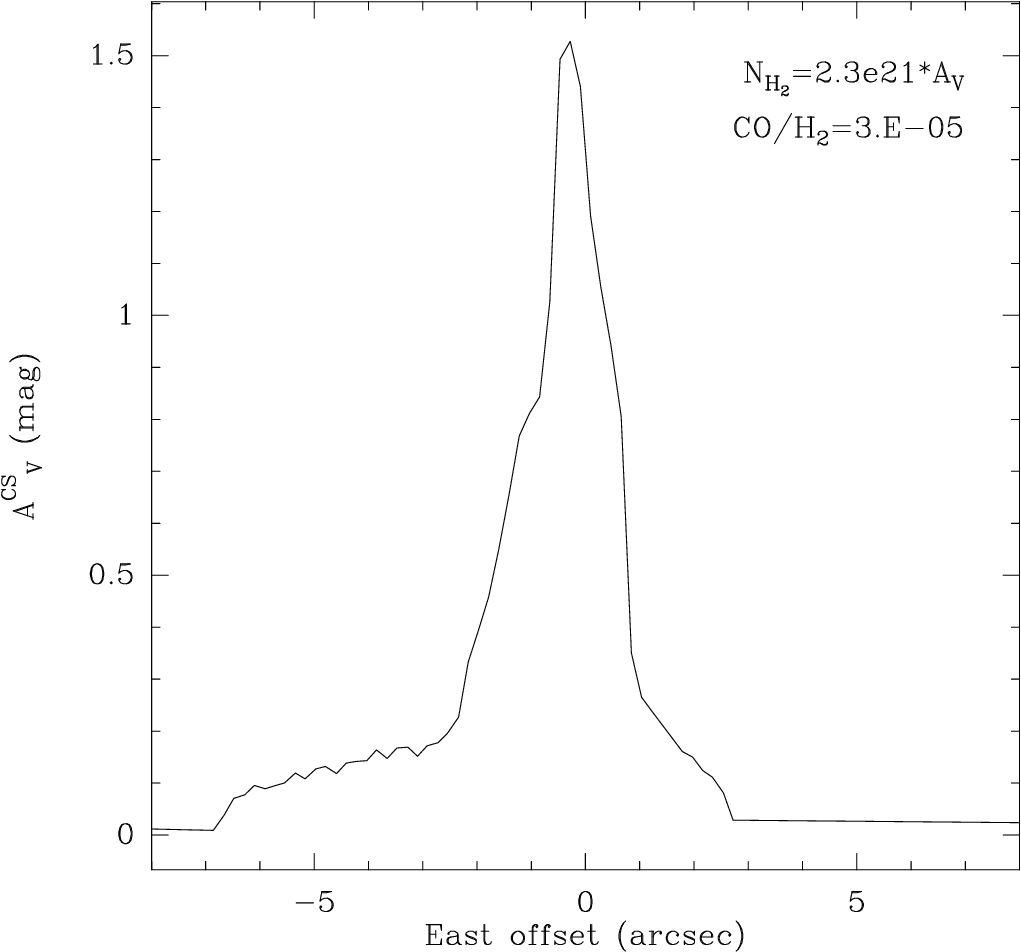}{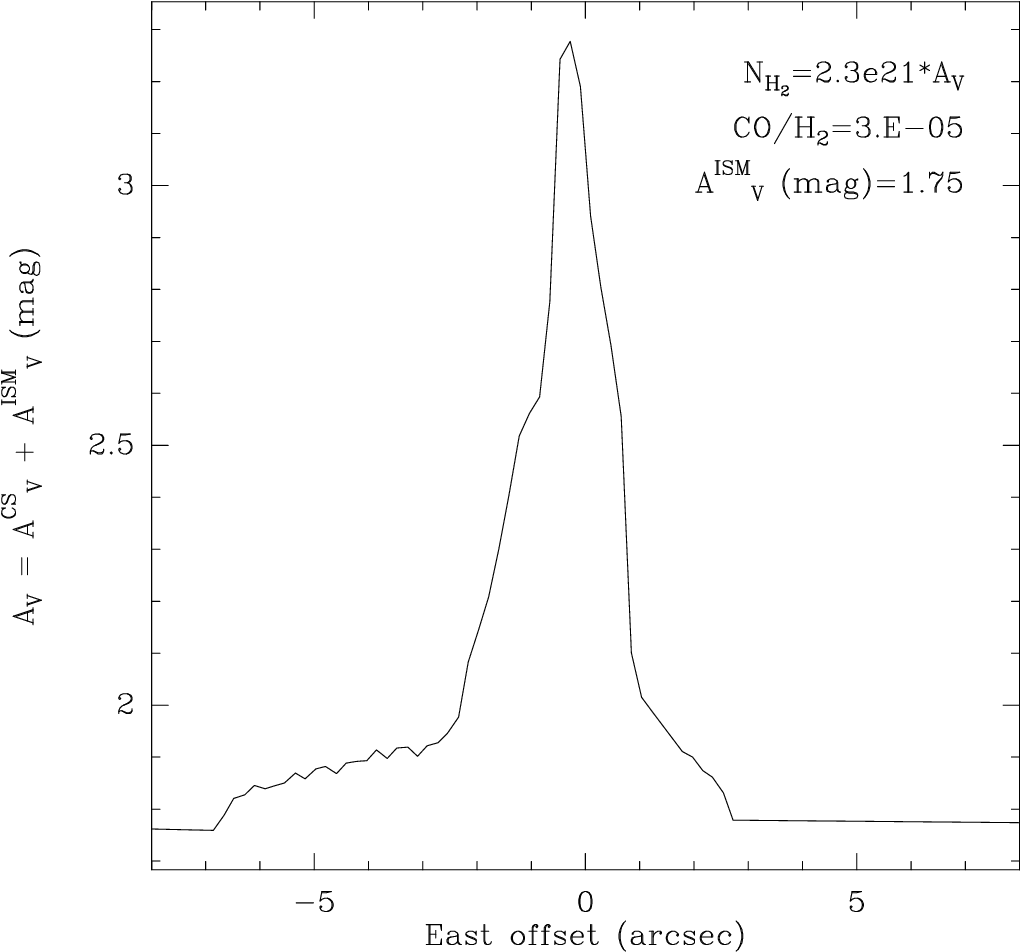}
\caption{Spatial distribution of the circumstellar (left) and total, i.e. circumstellar+interstellar, 
extinction (right) along PA90 inferred from the model of the molecular
envelope by \cite{cc02}. The value for the CO-to-H2 abundance ratio,
ISM extinction, and the conversion factor between H2 column density
and dust extinction adopted are indicated.}
\label{f_av}
\end{figure}

We have compared the extiction computed above with that predicted from
the detailed model of the molecular envelope of M\,2-56 performed by
\cite{cc02} based on high-angular resolution interferometric maps of
the CO\,($J$=1-0 and $J$=2-1) emission. The molecular envelope surrounds the optical 
nebula and is expected to contain a mixture of gas and dust. 
We have
derived the relative variation along the lobes of the CO column
density from the geometry and density spatial distribution
in the model. Adopting a given CO-to-H$_2$ relative abundance, X(CO), and
the standard conversion factor from H$_2$ column density to optical
extinction N$_{\rm H_2}$=2.3$\times$10$^{21}$\,A$_{V}$\,cm$^{-2}$, we have calculated the relative variation of the
circumstellar extinction (A$^{\rm CS}_{V}$) along the lobes produced by the dust in the
hourglass-shaped molecular envelope (Fig.\,\ref{f_av}, left). In order
to determine the total (circumstellar plus interstellar) extinction, we have added the
interstellar contribution in the direction to M\,2-56, which is A$^{\rm ISM}_{V}$=1.75 according
to the ``Galactic Dust Reddening and Extinction'' calculator provided
by IRSA/IPAC\footnote{\tt http://irsa.ipac.caltech.edu/applications/DUST/} (Fig.\,\ref{f_av}, right).

In their model, \cite{cc02} adopt a typical value of X(CO)=2$\times$10$^{-4}$, however, these authors affirm
that the average CO abundance in this object must be substantially lower 
due to significant photodissociation of this molecule. We find that
assuming X(CO)$\approx$3$\times$10$^{-5}$ we are indeed able to
reproduce very well the total extinction derived from the Balmer
decrement. (Note that most of the \hal\ and, especially, \hb\ emission
arise in the bright ILs, therefore our estimate of A$_{\rm
V}$$\sim$3\,mag represents the mean extinction over the central
$\pm$2\arcsec-3\arcsec).
We have obtained an independent estimate of X(CO) by comparing the
total dust mass in M\,2-56 obtained from simple SED modeling, $M_{\rm
dust}$=1.5$\times$10$^{-3}$\,\msun, with the mass of the molecular
envelope derived from CO, $M_{\rm gas}$=0.05\,\msun, assuming the
standard value of X(CO)=2$\times$10$^{-4}$. Both masses are related by
the following expression:
 
\begin{equation}
M_{\rm gas} = M_{\rm dust} \times [\frac{X(CO)}{2\times10^{-4}}] \times \delta,
\end{equation}

where $\delta$ is the gas-to-dust mass ratio, which is typically in
the range $\delta$=100-200 in pPNs.
Applying the previous equation, we deduce an average CO abundance of
X(CO)$\sim$(3.3-6.6)$\times$10$^{-5}$, comparable to that inferred from the
extinction and the CO model explained above. 


\subsection{A compact, dusty ``cocoon'' around the star?}
\label{cocoon}

We have estimated the total optical extinction along the line-of-sight
towards the central star of M\,2-56 (A$^{\star}_{V}$) from the ratio
between the continuum flux observed at a given wavelength ($F^{\rm obs}_{\lambda}$) and that expected from the central star in the
absence of absorption ($F_{\lambda}$).  We compute $F^{\rm
obs}_{\lambda}$ from the observed surface brightness at the nebula center and
taking into account the PSF in our observations.
On the other hand, $F_{\lambda}$ has been approximated by the flux
(measured on earth) emitted by a black body with an effective
temperature of \teff=25,000\,K, a total bolometric luminosity of
$L$=5500\,\ls, and a distance to M\,2-56 of $d$=2.1\,kpc
(\S\,\ref{intro}). Considering the values obtained for $F^{\rm
obs}_{6510}$=1.8$\times$10$^{-17}$\,erg\,s$^{-1}$\,cm$^{-2}$\,\AA$^{-1}$
and F$_{6510}$=4$\times$10$^{-13}$\,erg\,s$^{-1}$\,cm$^{-2}$\,\AA$^{-1}$, we
infer a value for the total extinction towards the star of A$^{\star}_{V}$=12\,mag.
The slope of the continuum emission over the whole
wavelength range observed by us in 1998 is indeed very well matched by
that predicted by the reddened black body above. 
The color of the continuum, however, is not compatible with a 
lower value of A$^{\star}_{V}$\,$\sim$A$_{V}$\,$\sim$3-4\,mag 
since, in that case, the central star would have
to be unreasonably hot (\teff$\ga$300,000 K), which is totally in
contradiction with the low-excitation conditions and lack of
substantial photoionization of the optical nebula of M\,2-56.

Our value of A$^{\star}_{V}$ is significantly larger than the average nebular
extinction implied by the Balmer decrement and also larger than
the maximum value deduced from the CO column density under a
reasonable assumption of the CO abundace (see \S\,\ref{extinction}). 
This result is not totally unexpected: the CO model only accounts for the extinction
produced by the extended hourglass-like envelope that surrounds the
optical lobes, where the \hal\ and \hb\ emission is locally produced,
but does not include the possible attenuation of the stellar radiation
produced by a potential/conceivable dusty structure {\it inside} the
optical lobes and closer to the stellar surface.
The absence of scattered continuum along the lobes in 1998 suggests
that the light from the star was largely attenuated in that epoch not
only along the line-of-sight, as indicated by the large
A$^{\star}_{V}$ found, but also along the nebular symmetry axis, i.e.\
in the direction to the lobes. It is then quite plausible that the
structure responsible for the large optical depth towards the star 
was completely (or almost completely) surrounding the latter 
in 1998 forming, for example, a dusty ''cocoon'' or shell. 
Since such a cocoon or shell-like obscuring structure is most likely
surrounding the close stellar environment and, thus, deep inside the
ILs, we derive an upper limit to its radius of $<$0\farcs1 ($<$200 AU)
from the PSF of our high-angular resolution $HST$ images.

%
%
\subsection{Recent decrease of the optical depth towards the star}
\label{cocoon_09}

The remarkable spectral differences observed in 2009 with respect to
1998 and 2000 (\S\,\ref{res_09}) signal important physical changes at
the core of M\,2-56. On the one hand, the recent appearance in 2009 of
scattered continuum all along the lobes suggests that: 1) the
intrinsic intensity of the stellar continuum has increased; and/or 2)
the optical depth of the material (cocoon?) around the star has
decreased (at least in the direction to the lobes) leading to a larger
fraction of the stellar radiation to escape and to reach the lobes
where it is reflected by the dust. This (option 2) would also explain
the increment of the scattered (versus locally produced) \hal\
emission along the lobes in 2009 and, in particular, the recent
brightening of clumps C3 and C3' evidenced by the direct images
(Fig.\,\ref{f_images}). On the other hand, we have observed a
significant brightening of the continuum between 1998 and 2009. In
particular, at the nebula center we measure $F^{\rm
obs}_{6510}$(2009)$\sim$15$\times$$F^{\rm obs}_{6510}$(1998). We
believe that a true variation of the luminosity or the stellar
temperature is very unlikely to be the cause of such
brightening. First, an increase of the luminosity would be totally
unexpected in the post-AGB phase, which is characterized by a
progressive warming of the central star at constant
luminosity. Second, only a decrease of the stellar temperature down
to \teff$\sim$5500-6000\,K (at constant luminosity L=5500\ls) combined
with an extinction of A$^{\star}_{V}$=11.7\,mag would be able to
reproduce the continuum spectrum observed in 2009. Such an extreme
cooling of the star, however, would not only be unprecedented but,
most importantly, would be inconsistent with a) the presence of a
compact circumstellar
\ion{H}{2} region in 2009, which requires a relatively warm star (\teff$\ga$20,000\,K), 
and b) the lack of metallic absorption lines typical of a G-type star
in the stellar spectrum of M\,2-56.

Accordingly, the increase of scattered (continuum and
\hal) emission along the lobes and the brightening of the continuum at the center 
are both most likely due to a decrease of the optical depth 
of the circumstellar dust enshrouding the nebula's core (which
includes the star and the compact circumstellar \ion{H}{2} region).

Applying the same method and adopting the same value for F$_{6510}$ as in \S\,\ref{cocoon} and
taking into account the continuum flux measured in 2009 near \hal,
$F^{\rm obs}_{6510}$=2.7$\times$10$^{-16}$\,erg\,s$^{-1}$\,cm$^{-2}$\,\AA$^{-1}$
(\S\,\ref{res_09}) we deduce a value of
A$^{\star}_{V}$=8.8\,mag. Therefore we conclude that the extinction towards the star 
along the line-of-sight has decreased by $\sim$3\,mag from 1998 to 2009. 

We have also estimated a lower limit to the decrease of the optical
depth from the star to the lobes, in particular, to the location
of condensation C3 of the eOL (Figs.\,\ref{f_images} and
\ref{f_2009all}) from the ratio of the surface brightness of the
scattered continuum at that point in 2009 and in 1998. The stellar
scattered continuum observed at C3 is attenuated by $i$) the dust inside
the lobes, i.e.\ from the star to C3 (symbolized as $\star
\rightarrow$ C3), and also by $ii$) the dust between C3 and us along the
line-of-sight, i.e.\ contained in the extended CO envelope that
surrounds the optical lobes and in the ISM. However, the extinction
produced by the latter components is not expected to vary with time
(at least, not in less than $\approx$10\,yr). In 1998, the continuum
is not detected, therefore we use an upper limit given by the rms
noise in our spectra
(1$\sigma$=3.1$\times$10$^{-18}$\,erg\,s$^{-1}$\,cm$^{-2}$\,\AA$^{-1}$\,pix$^{-1}$). In
2009, we measure
1.1$\times$10$^{-17}$\,erg\,s$^{-1}$\,cm$^{-2}$\,\AA$^{-1}$\,pix$^{-1}$
at 6510\AA.  Therefore, we conclude that the extinction from the star
in the direction to the lobes has decreased by A$^{\star
\rightarrow C3}_{V}$$>$1.6\,mag from 1998 to 2009.


\section{Density and Nebular Mass}
\label{nemass}

\subsection{The electron density distribution}
\label{ne}

\begin{figure}
\epsscale{0.265}
\rotatebox{270}{\plotone{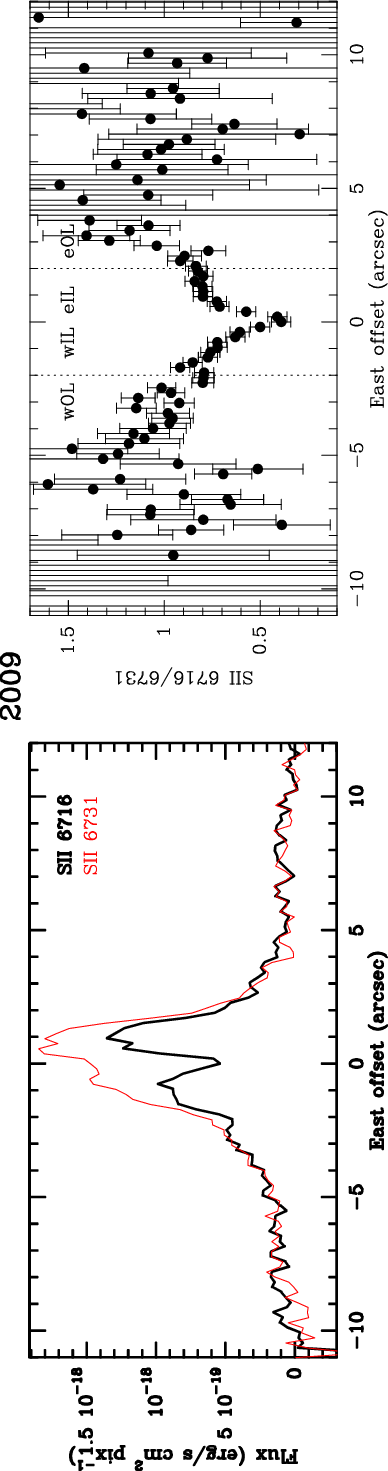}}
\begin{center}
\epsscale{0.45}
\rotatebox{270}{\plotone{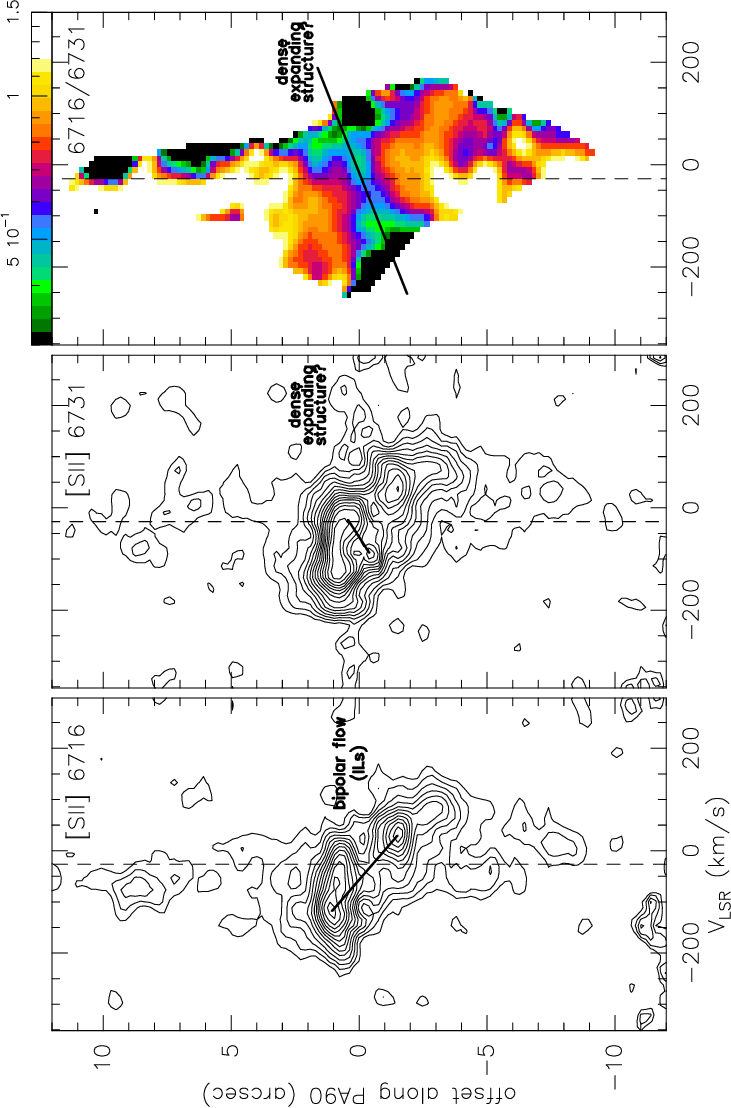}}
\end{center}
\caption{{\bf Top)} Spatial distribution along PA90 of the lines of the 
[\ion{S}{2}]$\lambda\lambda$6716,6732\AA\ doublet (left) and
their relative intensity (right) observed in 2009.  {\bf
Bottom)} Position-velocity diagram along PA90 of the [\ion{S}{2}] lines and their ratio 
observed in 2009. The p-v distribution of the
6716/6731 intensity ratio towards the center evidences a dense,
equatorially expanding structure.}
\label{f_ne}
\end{figure}

We have used the [\ion{S}{2}]$\lambda$6716/$\lambda$6731 doublet ratio
to estimate the electron density, $n_{\rm e}$, along the shocked lobes
of M\,2-56. In Fig$.$\,\ref{f_ne}, we plot the spatial profiles along
the nebula axis (slit PA90) for the two doublet lines and their ratio
obtained from the spectrum observed in 2009 (with the highest S/N
ratio and best spatial resolution amongst our multi-epoch data). We
have checked that the ratios measured in 1998 and 2000 (not shown
here) are roughly consistent with those obtained in 2009 within the
observational errors. 



The [\ion{S}{2}]$\lambda$6716/$\lambda$6731 ratio increases with the
axial distance from the nebula center, which implies a decrease of
$n_{\rm e}$ from the center to the outer parts of the lobes,
considering the [\ion{S}{2}]$\lambda$6716/$\lambda$6731 ratio versus
$n_{\rm e}$ relationship \citep[e.g.][]{ost06} and adopting an average
electron temperature $T_{\rm e}$\,=\,10,000\,K consistent with the estimate by 
\cite{coh77,tra93}.  We derive the largest value of $n_{\rm
e}$$>$2$\times$10$^4$\cm3\ towards the nebula center (within the PSF,
i.e.\ $\pm$0\farcs5), where the high-density limit of the
[\ion{S}{2}]$\lambda$6716/$\lambda$6731 ratio, $\sim$0.4, is reached.
Along the ILs, $n_{\rm e}$ ranges between 2000\,\cm3, at the base
($\pm$0\farcs5), and 1000\,\cm3, at the tips ($\pm$2\arcsec). In the
inner regions of the OLs (from $\pm$2\arcsec\ to $\pm$3\arcsec), the
density steeply varies between $\la$1000 and 500\,\cm3 and then
decreases down to 150\,\cm3 at $\pm$5\arcsec. Beyond these regions the OLs
may become more tenuous, if the spatially decreasing trend of $n_e$ is
maintained, however, the large data errorbars prevent an accurate
determination of the doublet ratio.


The p-v diagram of the two lines of the [\ion{S}{2}] doublet and their
intensity ratio, which are represented in Fig.\,\ref{f_ne}, provides
more details on the distribution of $n_e$ in M\,2-56. In particular,
unveils the presence of a dense equatorial region at the center
(approximately within $\pm$1\arcsec) with a velocity gradient markedly
different from (opposite to) that in the lobes. Such a velocity
gradient is indicative of expansive motions in the direction
perpendicular to the nebula symmetry axis. The velocity and the
density along the equatorial flow are both found to increase with the
distance to the center, reaching values of up to $\approx$\,100\,\kms\
and $\ga$2$\times$10$^4$\,\cm3. (We cannot rule out a radial decrease
of the electron temperature being totally or partially responsible for
the radial decrease of the [\ion{S}{2}]6716/6731 ratio).  The center
of the expansion may be slightly bluewards of
\vsys, at $V_{\rm r}$$\sim$$-$60\,\kms.


Finally, we have estimated a lower limit to the density of the
region/wind responsible for the transient feature F1 observed in the
\hal\ profile (and, to a lesser extent, the
[\ion{N}{2}]$\lambda\lambda$6548,6584\AA\ doublet) in 1998. We believe
that the most likely reason why feature F1 had vanished in 2000 is that the
material ejected during the \hal-emission burst had become fully (or
almost fully) recombined in the time span between our runs
\#1 and \#2, i.e.\ in just 2.9\,yr or less. This enables estimating a lower
limit to the average density in the fast F1-wind in 1998 of $n_{\rm
e}$$\ga$2.6$\times$10$^4$\cm3\ using the expression for the
recombination time scale, $t_r$, as a function of the density given by 
\cite{kwo00}: $t_r \sim 7.6 \times 10^4/[n_e/\cm3]$\,yr. Such a value of the density, relatively high, is
consistent with feature F1 not being present in the [\ion{S}{2}]
spectrum: given the critical density of the lines in the doublet these
transitions would be significantly suppressed by collisional
de-excitation in high-density regions. The lines of the [\ion{N}{2}]
doublet have larger critical densities, $n_{\rm
c}$$\sim$few$\times$10$^4$\,\cm3, that is why the F1-wind is traced by 
these transitions.

\subsection{Atomic and Ionized Mass}
\label{mass}

We have estimated the total mass of atomic and ionized gas, $M_{\rm
H}$ and $M_{{\rm H}^+}$, in the lobes of M\,2-56 using the mean
electron densities derived above and the total energy radiated by
[\ion{O}{1}]$\lambda$6300 and H$\alpha$, respectively.  The
[\ion{O}{1}] (H$\alpha$) intensity is proportional to the product of
$n_e$ and the H (H$^+$) number density, assuming that the transitions
are optically thin and that the electron temperature does not strongly
vary within the emitting region. Considering a mean electron
temperature in the shocked lobes of T$_{\rm e}$=10,000\,K
\citep{coh77,tra93}, relative abundances of He/H\,=\,0.1 and
O/H=\,4$\times$10$^{-4}$ 
and the required atomic parameters \citep[see e.g.][]{men83,gur97} we derive:

\begin{equation}
\label{mo1}
M_{\rm H}^{[\rm OI]}(\msun) = 7.7\times10^{-5} 
\left(\frac{n_e}{10^4 \cm3}\right)^{-1} 
\left(\frac{L_{[\rm OI]}}{0.1L_{\odot}}\right), \\ 
\end{equation}

\begin{equation}
\label{mha}
M_{{\rm H}^+}^{\rm H\alpha}(\msun) = 9.6\times10^{-5} 
\left(\frac{n_e}{10^4 \cm3}\right)^{-1} 
\left(\frac{L_{\rm H\alpha}}{0.1L_{\odot}}\right), 
\end{equation} \\

\noindent
where $L$ is the dereddened luminosity of the line (and we have
adopted $d$=2100\,pc). Note that Eq$.$ (\ref{mo1}) is valid for
electron densities smaller than the critical density of [\ion{O}{1}],
$n_{\rm c} \sim 10^6$\,\cm3, which is the case for M\,2-56. On the
other hand, this equation only provides a lower limit of the total
atomic mass ($M_{\rm H}$, which includes the He contribution), since
we are assuming that most of the oxygen is neutral and in the ground
state. Also, we note the dependence of the mass derived using Eq$.$
(\ref{mo1}) on T$_{\rm e}$ as $M_{\rm H}^{[\rm OI]} \propto
\sqrt(T_{\rm e})/(e^{-2.3e4/}T_{\rm e})$.
In deriving Eq$.$ (\ref{mha}) we
have assumed that the majority of the electrons in the nebula come
from H$^+$, i.e., $n_e \sim n_{H^+}$, and the classic radiative
recombination case b for $T_{\rm e}$\,$\sim$\,10,000\,K
\citep[i.e$.$ {\it z$_{\rm 3}$}=$n_3$/$n_{\rm e}n_{\rm
H^+}$=0.25$\times$10$^{-20}$\,cm$^3$, where $n_3$ is the population of
the level $n=3$ of H;][]{gur97}.

We have separately estimated the masses of the OLs and ILs using the
line fluxes integrated within each component and using Eqs$.$
(\ref{mo1}) and (\ref{mha}). We have also estimated $M_{{\rm H}^+}$ in
the F1-wind calculating the flux in the feature F1 computed as the difference
between the \hal\ fluxes measured in 1998 and 2000. To derive the
total mass in the extended OLs we have used the fluxes in Table
\ref{t_masses} multiplied by a factor $\sim$3, which converts the flux
within a 1\arcsec-wide slit to total nebular flux (estimated from the
direct $HST$ and ground-based images, Fig$.$\ref{f_images}). We
assume, however, that the compact ILs and wind-F1 lie completely
within our 1\arcsec-wide slit.  Our results are given in
Table\,\ref{t_masses} together with the average values of the
extinction and electron densities used for each nebular component
(estimated in \S\,\ref{extinction} and \S\,\ref{ne}). For the F1-wind our value of
the ionized mass is particularly uncertain since the density and the
extinction are poorly known in this case. The extinction by which the
flux in the F1 feature needs to be corrected for could be of up to
$\sim$12, if it is extinguished similarly to the central star, 
or $\sim$3, if it is extinguished similarly to the ILs. 
Given the angular size of the F1-emitting region (0\farcs5), which is
larger than that inferred for the cocoon around the nebula's core
($<$0\farcs1, \S\,\ref{cocoon}), we favor a moderate extinction (i.e.\ closer to $\sim$3) 
for this component. Given the lower limit to $n_e$, we calculate an upper limit to the mass of
the F1-wind of 
$<$4$\times$10$^{-6}$\,\msun. 

We have also estimated the mass of ionized material in M\,2-56
geometrically, using the mean electron densities and the volume of the
shocked lobes directly measured from the images, assuming a simplistic
spherical geometry for the lobes. We have separately
considered the case of hollow lobes and lobes uniformly filled with
gas.
Assuming hollow lobes with $\sim$0\farcs2-thick walls, the ionized
mass of the ILs and OLs is $\sim$7$\times$10$^{-5}$\,\msun\ and
$\sim$6$\times$10$^{-4}$\msun, respectively. These values are
consistent with those previously obtained from the H$\alpha$
luminosity (Table\,\ref{t_masses}). For the adopted values of the mean
electron density and taking into account the limitations of our necessarily
simplistic assumptions made to derive the mass, the
thin-walled lobes scenario seems to be a better approximation to the
structure of the lobes of M\,2-56 than the filled-lobes geometry: for
the latter, the derived masses systematically exceed those obtained
from the H$\alpha$ luminosity. We note that the limb-brightnening of
the ILs and, to a lesser extent, the OLs observed in the images
support the hypothesis of relatively thin-walled lobes. 

\section{Discussion: formation and evolution of M\,2-56}
\label{dis}

We interpret the nebular morphology and kinematics of M\,2-56 in the
context of the AGB-to-post-AGB wind interaction scenario, i.e., 
resulting from the hydrodynamical interaction between fast, collimated 
(jet-like) post-AGB ejections and the slowly expanding envelope expelled during
the previous AGB phase (see \S\,\ref{intro}). 
The remarkable ``lobe inside a lobe'' appearance of M\,2-56 together
with the discovery of the recent \hal\ emission burst, and the
revelation of other very small-scale structures at the nebula's core
points to multiple episodes of mass outflow during the post-AGB
phase. 

The earliest post-AGB mass-loss event in M\,2-56 (that we have
knowledge of from existing data) is probably the one that shaped and
accelerated the extended hourglass-like molecular outflow enclosing
the optical nebula (\S\,\ref{intro}). As discussed by \cite{cc02},
such an interaction started approximately $\sim$1400\,yr ago and took
place quickly, in less than 300\,yr \citep[see also][]{buj01}.
The smaller kinematical ages of the optical OLs and ILs derived from
our long-slit spectra, $t_{\rm kin}$$\sim$380 and $\sim$40\,yr,
respectively, suggest that the latter resulted from two consecutive
and more recent post-AGB ejections. The duration of such ejections
cannot be asserted based on our data but it may be a small fraction of
their kinematical ages as in the case of the fast molecular
outflow. We find that the caps of the optical OLs, which are located
inmediately behind of two diametrally oposed dense axial clumps traced
by CO, show a different kinematics than the main body of the OLs. In
particular, the OLs-caps are characterized by expansion velocities
close to those of the aforementioned CO condensations. The kinematical
age derived for the OLs-caps, $t_{\rm kin}$$\sim$1200\,yr, is thus
similar to that of the CO axial clumps, which may suggest that both
were shaped and accelerated in the same wind interaction process
between the primer post-AGB bipolar flow(s) and the AGB envelope.
In this scenario, the massive molecular envelope would be mainly composed of 
shocked-AGB wind whereas the OLs-caps could be a remnant of the primer 
post-AGB wind.  
Alternatively, the material in the OLs-caps
could have been ejected at the same time as the rest of the OLs, i.e.\
$\sim$380\,yr ago, and subsequently decelerated (from expansion speeds
of $\sim$400\kms\ down to the observed velocities in these regions,
\vexp$\sim$110\kms) as a consequence of its interaction with the pre-existing
CO clumps.
The fast ILs probably trace a more recent shock interaction 
between an ensuing, fast post-AGB wind with the material at the base of the pre-existing
and slower OLs.

Our data also uncover a dense equatorial structure at the nebula center
that is expanding in the
direction perpendicular to the lobes with radial velocities of up to
$\sim$100\,\kms\ (\S\,\ref{ne} and Fig.\,\ref{f_ne}). 
The velocity gradient observed, which is approximately linear,
indicates a kinematical age of $\sim$300-400\,yr (adopting
$i$=90\degr$-$18\degr=72\degr). The similar age of the equatorial
flow and the OLs is consistent with both components resulting from a
single, sudden mass-ejection event that took place simultaneously
along the nebula axis and in the perpendicular plane. This scenario
has been proposed previously by \cite{alc07} to explain the radial
acceleration along the equator and bipolar flows of the molecular
envelope of the pPN M\,1-92.
In M\,2-56, the compact equatorial structure unveiled by the [\ion{S}{2}]
doublet may represent the inner (warmer) regions of the dense,
molecular ring/torus probed by CO emission (\S\,\ref{intro}). The
lower expansion velocity of the latter (\vexp=7-8\,\kms) may indicate
substantial deceleration of the equatorial flow after interaction with
the dense, slow AGB envelope. 
The density is found to increase outwards along the optical equatorial
flow, in support of the presence of compressed (swept up) material at
the boundary with the dense molecular torus.
%


The discovery of the \hal\ emission burst from the nebular core in
1998 (feature F1) brings to light a latter stellar wind. As discussed
in \S\,\ref{f1wind}, the F1-emitting wind is probably bipolar.  The
linear size and mean expansion velocity of this fast, compact flow
allow us to derive an upper limit to its kinematical age of only
$<$14\,yr. The upper limit arises because of possible deceleration of
the F1-wind by interation with the ILs. The absence of feature F1 in
the \hal\ spectrum obtained in 1989 by G91, in fact, suggests that the
F1 wind could have been ejected only $<$10\,yr before it was first
identified in our run\,\#1. In this case, the initial expansion speed of
the F1-wind (before substantial deceleration took place) should have
been
\vexp$\sim$500\kms\ to explain the axial extent of this component. This value of \vexp\ 
is indeed in agreement with that inferred from the broad wings and the
P-cygni absorption component of feature F1, which indicate that a
fraction of the F1-flow is actually moving at this relatively large
speeds (\S\,\ref{f1wind}).

The absence of feature F1 in 2000 is probably due to fast
recombination of the gas in the dense ($>$2.6$\times$10$^4$\,\cm3)
F1-wind (\S\,\ref{ne}). In order for the F1-wind to be completely or
considerably neutral already in 2000, most of it
should have been far away (detached) from the central star 
at that time, otherwise one would expect to observe emission from the
dense photoionized material at the base of the wind close to the
stellar surface as long as the wind is on-going. We conclude, therefore, that the rapid recombination
observed, in less than $\sim$3\,yr, is most consistent with the
F1-wind being a short-time duration event, e.g.\ a pulsed-jet or
bullet-type ejection, leading to detached or clumpy structures, rather
than being blown in a continuous way. 
We believe that the clumpy, jet-like structures observed in the direct
$HST$ images of M\,2-56 at the base of the eIL could represent the
footprints of such a fast F1-wind. This is based on the similar
size and location of both structures in the images and the spectra. If
this is correct, the emission that we see from the prominent
three-finger like structure in the F606W $HST$ image must be mainly
stellar light $scattered$ by the F1-wind, after recombination, since
that image was obtained in 2002 once the \hal\ emission feature F1
had dissapeared.



We have estimated a lower limit to the mass-loss rate that led to the
main nebular components identified in our optical imaging and
spectroscopic data, namely, the OLs, the ILs and the F1-wind, from the
ratio between the total mass contained in each component and their
kinematical ages (Table\,\ref{t_masses}). The lower limit arises
because the duration of each of the multiple post-AGB mass-loss episodes 
undergone by M\,2-56 could be shorter than their kinematical ages. 
We obtain very similar values of the mass-loss
rate for the OLs and ILs, namely, \mloss$_{\rm OLs}$$\sim$2$\times$10$^{-6}$ and \mloss$_{\rm
ILs}$$\sim$3$\times$10$^{-6}$\,\my. For the F1-wind, we compute a
value of \mloss$_{\rm F1}$$\sim$4$\times$10$^{-7}$\,\my, however, we
note the larger uncertainties in this case mainly due to our vague
estimate of the mass of the F1-wind. We have attempted a different
estimate of the mass-loss rate of the F1-wind as given by 
$\mloss=\pi r_j^2 \vexp \rho$, 
where the geometry of the F1-wind is approximated by a cylindrical structure
with radius $r_j$, density $\rho$, and expanding at \vexp. 
Adopting $r_j$$\sim$0\farcs05, $\rho$$>$3$\times$10$^4$\,\cm3 and \vexp$\sim$500\,\kms, we derive 
\mloss$_{\rm F1}$$\sim$4$\times$10$^{-7}$\,\my. The radius adopted for the F1-wind 
is that of the bright, jet/finger-like features in the F606W $HST$
images that may be footprints of the F1-wind. The agreement between
our two estimates of \mloss$_{\rm F1}$ suggests that the order of
magnitude obtained may be correct.

In general, very little is known about mass-loss rates of post-AGB
winds, their continuous or episodic nature, and about the structure of the resulting
flows after the AGB-to-'post-AGB' wind interaction.  Our data indicate
that the various post-AGB mass-loss episodes experienced by M\,2-56
did not happen at regular time intervals, in particular, the time span
between two consecutive post-AGB ejections has shortened with
time. Also, according to the different expansion velocities of the
distinct nebular components of M\,2-56, including the molecular
envelope, the succesive post-AGB winds seem to be characterized by
ejection speeds exponentially increasing with time. 
It is also worth mentioning that, unlike most
pPNs and PNs with multiple optical lobes, the OLs and the ILs of
M\,2-56 are oriented almost identically (with their main symmetry axis
inclined with respect to the line-of-sight by $i$=18\degr;
\S\,\ref{inc}). Since the molecular envelope has also a similar
orientation, we conclude that the different post-AGB bipolar ejections
in this object (except maybe for the F1-wind) have taken place along
the same (or nearly the same) direction/axis over the last
$\sim$1400\,yr. We note that, in contrast, signatures of significant
jet directional changes are common amongst pPNs and PNs
\cite[e.g.][and references therein]{mir99,sahs07}. Another example of
directionally stable jets is found in the PN He\,2-90 \citep{gue01}.

We have obtained a rough estimate of the scalar linear momentum
carried by the bipolar OLs, the ILs, and the F1-wind from the product
of their masses (Table\,\ref{t_masses}) by their maximum expansion
velocities, which have been taken to be $\sim$250, 350, and 500\,\kms,
respectively. In this calculation we assume that the nebular
components are elongated structures with a velocity field that is
mainly axial (\S\,\ref{model}). The values found ($P_{\rm
OLs}$$\sim$0.17\,\msun\kms\ $>$ $P_{\rm ILs}$$\sim$0.05\,\msun\kms\
$>$ $P_{\rm F1}$$\sim$2$\times$10$^{-3}$\,\msun\kms) are orders of
magnitude smaller than the momentum carried by the molecular outflow,
$P_{\rm mol}$$\approx$10\,\msun\kms\ (in deriving this value, the
linear momentum computed by Castro-Carrizo et al., 2002, $P_{\rm
mol}$$\sim$2\,\msun\kms, has been corrected for our lower estimate of
the CO-to-H$_2$ molecular abundance -- see
\S\,\ref{extinction}).
According to this, we must conclude that none of the post-AGB ejections
probed by our optical data could have transferred its large linear
momentum to the CO outflow and, therefore, a primer more energetic
post-AGB ejection is necessary to explain the dynamics of the molecular
envelope. This result is in good accordance with the time decreasing
trend of the linear momentum of the optical nebular components
inferred by us ($P_{\rm OLs}$$>$$P_{\rm ILs}$$>$$P_{\rm F1}$), which
independently suggest (by extrapolation) that earlier post-AGB flows could
have carried a larger linear momentum.

The presence of a compact dusty structure and a nuclear \ion{H}{2}
region unveiled by our data (\S\,\ref{cocoon} and \ref{res_hii}) show
evidence of further post-AGB mass-loss after the F1-wind in
M\,2-56. The decrease of the optical depth along the line-of-sight
and along the nebular symmetry axis observed (\S\,\ref{cocoon_09}) is
consistent with a detached/shell-like (cocoon?) structure moving away
from the central star. Assuming that the cocoon expands at constant
velocity, $v_{\rm c}$, and that the density varies with the radial
distance to the star as $\rho \propto r^{-2}$ (i.e.\ the mass-loss
rate is constant), it can be easily demonstrated that for a thick and
a thin shell-like geometry:

\begin{eqnarray}
v_{\rm c}=r_{98}\times \left(\sqrt{\frac{N_{98}}{N_{09}}}-1\right) / \Delta t \hspace{1cm} {\rm (thick\ shell)} \\
v_{\rm c}=r_{98}\times \left(\frac{N_{98}}{N_{09}}-1\right) / \Delta t  \hspace{1cm} {\rm (thin\ shell)}, 
\end{eqnarray}

\noindent 
where $r_{98}$ is the inner and mean radius of the thick and thin cocoon, respectively, 
in 1998, $N_{98}$ ($N_{09}$) is the column density in 1998 (2009), and
$\Delta t$ is the time scale of the variation of the optical depth.
In 1998 and 2009, the total extinction towards the star was
A$^{\star}_{V}$=12\,mag and 8.8\,mag, respectively
(\S\,\ref{cocoon_09}). Subtracting the component of the extinction
produced by the CO envelope and the ISM at the center, $\sim$3\,mag,
we estimate that the $V$-band extinction produced by the cocoon was 
A$^{\rm cocoon}_{V}$=12-3=9\,mag in 1998 and 
A$^{\rm cocoon}_{V}$=8.8-3=5.8\,mag in 2009. 
Since the extinction is proportional to the column density, we find
$N_{98}/N_{09}$=9/5.8=1.55.  Adopting $r_{98}$$\sim$0\farcs1
($\sim$3.14$\times$10$^{15}$\,cm, \S\,\ref{cocoon}) and $\Delta t$$\sim$10\,yr, the variation
of the optical depth observed can be explained for moderate expansion
velocities of the cocoon of $v_{\rm c}$$\sim$30 and 60\,\kms\ for the
thin and thick shell approximation, respectively. 

In the previous expanding cocoon scenario, the detached geometry
(needed to explain the decrease of the optical depth) implies that the
mass-loss process that led to this structure had already ended in
2009. The compact size of the cocoon, which is well inside the ILs and
plausibly closer to the star than the F1-wind, suggest it is at least
as young as the latter (i.e., $t_{\rm kin}$$<$10\,yr). Adopting an
expansion velocity of $v_{\rm c}$=500\,\kms\ as for the F1-wind, which
is not unreasonable given the increasing trend of the expansion
velocity of the multiple post-AGB winds of M\,2-56, we obtain $t_{\rm
kin}$$\la$\,2\,yr.

The presence of compact cocoons enshrouding the central source has
been suggested in several pPNs. For example, the so-called {\sl
searchlight beam} features that are observed emerging from the nebular
core in an increasing number of pPNs and PNs (of which CRL\,2688
remains as the best known example) have been hypothesized to result
from starlight escaping through holes/cavities in an inner dust cocoon
\citep{sah98,san07}. Since these cocoons have not been 
{\sl directly} detected, their existence remains speculative so far.
We cannot rule out that the large obscuration of the
central source of M\,2-56 both along the lobes and the equatorial
plane could also be produced by {\sl two} distinct nebular components,
namely, a dusty bipolar flow and an equatorial disk/torus, rather than
by a unique cocoon-like structure. Direct observations of the
obscured, innermost regions of M\,2-56 using, for example,
high-angular resolution techniques in the infrarred and mm-wavelength range are 
crucial to characterize the circumstellar geometry at the nebula's heart. 

Finally, our spectra taken in 2009 have revealed a compact
\ion{H}{2} region around the central star of M\,2-56 that may represent the 
very latest (current?) mass-loss episode of this object.  It is
possible that such a compact \ion{H}{2} region was already formed in
1998 but its light, like the stellar radiation, was heavily blocked
from our view by the dusty cocoon. Only due to the recent decrease of
the optical depth of the latter, the emission from the \ion{H}{2}
region has been able to reach the lobes where it is scattered by the
nebular dust, enabling its indirect detection. Alternatively, it is
also possible that the \ion{H}{2} region has formed recently (after
2000). In the following, we discuss this possibility in some
detail. The recent formation of the \ion{H}{2} region would imply an
equally recent and rapid evolution (in less than $\sim$10\,yr) of the
central star towards higher effective temperatures, needed to explain
the raise of photonionizing UV radiation and detectable \hal\ emission
from the ionized regions. Although a large increase of \teff\ is not
deduced from our multi-epoch spectroscopic data, which do not show
significant changes in the stellar spectrum\footnote{except for its
brightening as a result of the decrease of the cocoon optical depth --
see \S\,\ref{extinction}.}, a low/moderate increase of the stellar
temperature from a B\,0.5 to a B\,0 spectral type
($\Delta$\teff$\sim$3000\,K) is simultaneously consistent with a
stable B-type classification and with a large increase of the \hal\
equivalent width from $W_\lambda$=50 to 300\AA\ \citep[for details,
see][]{san08}.
The \teff\ evolutionary rate value deduced in this case,
$\Delta$\teff/$\Delta$t$\ga$3000/10$\sim$300\,K\,yr$^{-1}$, is larger
than that expected for the central star of M\,2-56 based on
theoretical evolutionary models for post-AGB objects taking into account
the low luminosity and, thus, initial mass of this object 
\citep[e.g.][]{blo95,hoof97}. However, 
these models assume constant values for the post-AGB mass-loss rate of
10$^{-7}$-10$^{-8}$\my, single-star evolution scenarios, and other
simplifying hypothesis that may not be appropriate for M\,2-56. In
particular, we do not rule out that the lattest sudden mass ejections
leading to the F1-wind and the compact cocoon, could have sped up/boost  
the evolution of the central star towards slightly higher \teff\ and the recent
emergence of a nuclear \ion{H}{2} region in just a few years.

In addition to the central \ion{H}{2} region, there are probably
shocks currently happening at the innermost circumstellar regions of
this fascinating object as evidenced by the recent emergence of
\ion{Fe}{2} lines from the nucleus
(\S\,\ref{res_09}). These \ion{Fe}{2} lines are known to be good
tracers of astrophysical shocks, e.g., in supernovae, Herbig Haro objects,
PNe, etc \citep[e.g.][]{wel99,rei00}.
The presence of shocks at the nebula's core 
suggests current, fast stellar ejections
(in agreement with the broad nuclear \hal\ profile)
interacting hydrodynamically with the material in the close
environment of the star ejected in earlier mass-loss episodes with
smaller velocities. 
High-angular resolution optical imaging and spectroscopy are needed
to properly study the spatio-kinematic structure of the innermost 
nebular regions and to attempt understanding the complex
mass-loss history of the latest post-AGB ejections in M\,2-56.

%

\section{Summary}
\label{summ}

We report multi-epoch long-slit spectra at various slit positions and
direct images in the optical of the pPN M\,2-56. These data have
allowed us to describe the spatio-kinematic structure and complex
(currently active?) mass-loss history of this object with unprecedented
detail.

\begin{itemize} 

\item[-] Our data probe several nebular components, namely, 
two pairs of nested, co-axial lobes with different sizes and a large
contrast in surface brightness referred to as the faint outer lobes
(OLs) and the bright inner lobes (ILs), a compact bipolar flow (the
F1-wind), and an equatorially expanding central structure. A compact
circumstellar structure obscuring the star (cocoon?) and a nuclear
\ion{H}{2} region, both spatially unresolved, are also inferred from
our data.

\item[-] The optical lobes are oriented along the East-West direction (PA=90\degr) and their 
symmetry axis is inclined $\sim$18\degr\ (with respect to the plane of
the sky), similarly to the hourglass-shaped molecular envelope that surrounds 
the optical nebula \citep{cc02}.

\item[-] The OLs and the ILs have an overall axial symmetry 
and are characterized by an expansive kinematics with the velocity
increasing with the distance to the nebula center, except for the 
outermost OLs-caps, which show reduced speeds relative to their 
innermost regions. The maximum velocity reached by the OLs (ILs) 
is $\sim$250\,\kms\ ($\sim$350\,\kms). Consistent with the large speeds
observed at the tips of the ILs, we measure proper motions ($\Delta \theta_{\rm t}$$\sim$0\farcs03 yr$^{-1}$) 
by comparing two-epoch $HST$ images.

\item[-] In 1998, we detected a burst of \hal\ emission from the nebula nucleus 
(referred to as ``feature F1'') that has vanished in less than 2.9\,yr.  
We believe that feature F1 arises in a dense, 
fast ($\sim$350-500\,\kms) bipolar
wind ejected after 1989 as a short-time mass-loss event. 
 
\item[-] The  mass in the OLs, ILs, and F1-wind is 
$M_{\rm OLs}$$\sim$7$\times$$10^{-4}$\,\msun, $M_{\rm
ILs}$$\sim$1.2$\times$$10^{-4}$\,\msun, and $M_{\rm
F1}$$\la$4$\times$$10^{-6}$\,\msun, respectively. 

\item[-] 
The p-v distribution of the [\ion{S}{2}]6716/6731 line ratio has led
to the discovery of a dense ($n_e$$\ga$10$^4$\,\cm3), equatorial
flow. The radial velocity increases approximately linearly with the distance
to the center, reaching values of up to $\sim$100\,\kms. It is
possible that the equatorial flow and the OLs both resulted from a
single, sudden mass-loss event that took place simultaneously along
the axis and in the perpendicular plane.

\item[-] 
We infer the presence of a compact ($<$200\,AU), dusty structure
(cocoon?)  enshrouding the central star. The optical depth of this
structure has decreased a few magnitudes both along the line-of-sight
and along the optical lobes in $\la$10\,yr.
This has enabled a fraction of the light from the
nebula's core to escape the stellar vicinity and reach the lobes where
it is scattered off by nebular dust, leading to a brightening of the
scattered stellar continuum and the increase of scattered \hal\
emission along the lobes in 2009.

\item[-] The scattered \hal\ emission detected in 2009 along the lobes arises most likely 
from a compact \ion{H}{2} region around the star. 

\item[-] In 2009, we have discovered a number of permited emission lines 
arising from the nucleus. The detection of \ion{Fe}{2} lines suggests
the presence of shocks at the stellar neighborhood resulting from
current stellar wind activity.

\item[-] The kinematical ages of the main nebular components of M\,2-56 
are different from each other: $t_{\rm k}$$\sim$1400\,yr for the
molecular bipolar flow, $t_{\rm k}$$\sim$300-400\,yr for the OLs and
the equatorial flow, $t_{\rm k}$$\sim$40\,yr for the ILs, $t_{\rm
k}$$\la$10\,yr for the F1-wind, and even smaller for the compact
cocoon and \ion{H}{2} region.

\item[-] The rapid evolution of M\,2-56 is driven by multipe episodes of mass
ejection, that is, through a gusty or episodic post-AGB wind, that has
led to the nested bipolar morphology of the nebula and the younger
nuclear components discovered in this work. The duration of each of
such ejections may be a small fraction of its kinematical age, which
implies extremely short life-times, of a few years or less, for some
of the mass-loss episodes in M\,2-56.

\item[-] The various post-AGB mass ejection events experienced by M\,2-56
did not happen at regular time intervals, in particular, the time span
between two consecutive post-AGB ejections has shortened with time.



\item[-] The successive multiple post-AGB winds in M\,2-56 are characterized by
ejection speeds increasing with time from 200 to $\ga$\,500\,\kms. 
In contrast, the
mass-loss rate and linear momentum may show a time decreasing trend. 
None of the post-AGB ejections probed by our optical data could have
transferred its large linear momentum to the molecular outflow
($P_{\rm mol}$$\approx$10\,\msun\kms) and, therefore, a primer more
energetic post-AGB ejection is necessary to explain the dynamics of
the molecular envelope.

\end{itemize}

In summary, the data presented here indicate that we are witnessing
the current, variable wind activity at the core of M\,2-56 and the
rapid evolution of the shocked nebular material in very short time
scales. This makes of M\,2-56 a unique object for studying the yet
poorly known processes responsible for nebular post-AGB evolution
through follow up studies. 
New hydrodynamical symulations of the AGB-to-'post-AGB' wind
interaction using input parameters for the post-AGB wind in accordance
with observational properties deduced from this work may be very 
useful for improving our understanding of PN/pPN shaping and
evolution. For example, post-AGB winds in the form of a series of
sudden, short-lived ejection events (not necessarily regularly spaced
in time) should be considered. Also, models should allow for time
variations not only in the velocity of the fast wind but also in its
mass-loss rate and linear momentum.



\acknowledgments
We are grateful to the anonymous referee for his/her valuable
comments. We thank Jorge Garc\ai a Rojas for performing the service
mode observations of M\,2-56 with the INT+WFC and NOT+ALFOSC. The INT
is operated on the island of La Palma by the Isaac Newton Group in the
Spanish Observatorio del Roque de los Muchachos of the Instituto de
Astrof\ai sica de Canarias. Part of the data presented here has been
taken using ALFOSC, which is owned by the Instituto de Astrof\ai sica
de Andaluc\ai a (IAA) and operated at the Nordic Optical Telescope
under agreement between IAA and the NBIfAFG of the Astronomical
Observatory of Copenhagen. This work has been partially performed at
Laboratorio de Astrof\ai sica Espacial y F\ai sica Fundamental
(LAEFF-CAB) and has been partially supported by the Spanish MICINN
through grants AYA\,2006-14876, AYA2009-07304, and CONSOLIDER INGENIO
2010 for the team ``Molecular Astrophysics: The Herschel and Alma Era
-- ASTROMOL'' (ref.: CSD2009-00038), and by DGU of the CM under
IV-PRICIT project S-0505/ESP-0237 (ASTROCAM). LFM is supported
partially by grant AYA2008-01934 of the Spanich MICINN (co-funded by
FEDER funds), grant FGM-1747 of the Junta de Andaluc\ai a, and grants
INCITE09E1R312096ES and INCITE09312191PR of the galician INCITE
research programme of the Direccion Xeral de Investigaci\'on,
Desenvolvemento e Innovaci\'on of the Spanish Xunta de Galicia.  This
research has made use of the SIMBAD database, operated at CDS,
Strasbourg, France, and NASA's Astrophysics Data System.



{\it Facilities:} \facility{INT} \facility{NOT}

\clearpage 

\clearpage
\begin{table}
\begin{center}
\footnotesize
\caption{Log of spectroscopic observations \label{t_log}}
\begin{tabular}{ccccccccc}
\hline             
\hline             
Run & Date & Telescope & Grating & Dispersion & Range & Pixel & Slit &
Exposure \\
\#  & (yyyy-mm-dd) & +Instrument  &  & (\AA\,pix$^{-1}$) &  (\AA)     & (\arcsec) & Position\tablenotemark{a} & (s)       \\
\hline             
1     & 1998-01-17   & INT+IDS  & R1200Y  & 0.39   & 6370-6770  & 0.33      & PA90     & 3600      \\
1     & 1998-01-18   & INT+IDS  & R1200Y  & 0.39   & 6370-6770  & 0.33      & PA0,PA0E & 3600,3600 \\
&&&&&&&&\\
2     & 2000-11-11\tablenotemark{b}   & INT+IDS  & R1200Y  & 0.22  & 6234-6805  & 0.19      & PA90,PA0E     & 10800,9000 \\
2     & 2000-11-12   & INT+IDS  & R1200Y  & 0.22     & 6234-6805  & 0.19      & PA0W,PA119 &  11700,10800 \\
2     & 2000-11-13   & INT+IDS  & R900V  & 0.30     & 4558-5330  & 0.19      & PA90     &     27300     \\
&&&&&&&&\\
3     & 2009-08-23   & NOT+ALFOSC   & VPH\,\#17 & 0.26 & 6330-6850 & 0.19 & PA90 & 6800 \\ 
\hline
\tablenotetext{a}{See Fig.\,\ref{f_spec2} for slit position labels.}
\tablenotetext{b}{Non-photometric conditions}
\end{tabular}
\end{center}
\end{table}

\input{t_fluxes.tex}  

\input{t_masses.tex}

\end{document}

%% file: t_fluxes.tex
\begin{deluxetable}{lrrrrrr} 
\tablecolumns{7} 
\tablewidth{0pc} 
\tablecaption{Multi-epoch line fluxes observed ($F_{\lambda}$) and derredened ($I_{\lambda}$) for M\,2-56} 
\tablehead{ 
\colhead{} &  \multicolumn{3}{c}{$F_{\lambda}$} & \multicolumn{3}{c}{$I_{\lambda}$} \\ 
\colhead{} &  \multicolumn{3}{c}{(10$^{-15}$ erg\,s$^{-1}$\,cm$^{-2}$)} & \multicolumn{3}{c}{(10$^{-15}$ erg\,s$^{-1}$\,cm$^{-2}$)}\\
\cline{2-4} \cline{5-7} \\ 
\colhead{Line} & \colhead{1998} & \colhead{2000} & \colhead{2009} & \colhead{1998} & \colhead{2000} & \colhead{2009}} 
\startdata
H$\beta$\,$\lambda$4861.3                       & \nodata 	 & 1.20(0.04) & \nodata    &\nodata	&36(9)	     &\nodata \\
$[$\ion{O}{3}$]$$\lambda$5006.8  		& \nodata        & 0.07(0.04) & \nodata    &\nodata	&2.0(1.2)    &\nodata \\
$[$\ion{Fe}{2}$]$$\lambda\lambda$5158.0,5158.8 	& \nodata        & 0.20(0.04) & \nodata    &\nodata	&4.9(1.4)    &\nodata \\
$[$\ion{N}{1}$]$$\lambda$5200.3 		& \nodata        & 0.65(0.04) & \nodata    &\nodata	&15(3)	     &\nodata \\
$[$\ion{Fe}{2}$]$$\lambda$5261.6 		& \nodata        & 0.05(0.04) & \nodata    &\nodata	&1.1(0.9)    &\nodata \\
$[$\ion{O}{1}$]$$\lambda$6300.3 		& \nodata 	 & 6.4(0.4)   & \nodata    &\nodata	&76(12)      &\nodata \\
$[$\ion{O}{1}$]$$\lambda$6363.8 		& \nodata        & 2.0(0.4)   & 2.2(0.14)  &\nodata	&23(6)       &25(4) \\
\ion{Fe}{2}$\lambda\lambda$6383.8,6385.5\tablenotemark{\dag}&\nodata &\nodata & 1.12(0.07) &\nodata 	&\nodata     &13(2) \\
\ion{Fe}{2}$\lambda$6416.9\tablenotemark{\dag}  & \nodata        & \nodata    & 0.21(0.07) &\nodata	&\nodata     &2.4(0.9) \\
\ion{Fe}{2}$\lambda$6432.7\tablenotemark{\dag}  & \nodata        & \nodata    & 0.42(0.07) &\nodata	&\nodata     &4.7(1.1) \\
\ion{Fe}{2}$\lambda$6456.4\tablenotemark{\dag}  & \nodata        & \nodata    & 0.77(0.07) &\nodata	&\nodata     &8.5(0.14) \\
\ion{Co}{1}?$\lambda$6490.3\tablenotemark{\dag} & \nodata        & \nodata    & 0.21(0.07) &\nodata	&\nodata     &2.3(0.8) \\
\ion{Fe}{2}$\lambda$6516.1\tablenotemark{\dag}  & \nodata        & \nodata    & 0.91(0.07) &\nodata	&\nodata     &10(2) \\
$[$\ion{N}{2}$]$$\lambda$6548.0 		& 2.3(0.2)  	 & 1.6(0.4)   & 4.9(1.1)   &24(4)	&17(5)       &52(14) \\
H$\alpha$\,$\lambda$6562.8                      & 23(0.4)   	 & 15.1(0.4)  & 161(1.1)   &240(40)	&160(6)      &1700(300) \\
$[$\ion{N}{2}$]$$\lambda$6583.5 		& 6.7(0.3)  	 & 4.8(0.4)   & 10.5(1.1)  &70(11)	&50(2)	     &110(20) \\
\ion{Fe}{1}$\lambda$6663.4\tablenotemark{\dag}  & \nodata        & \nodata    & 0.14(0.07) &\nodata	&\nodata     &1.4(0.2) \\
$[$\ion{S}{2}$]$$\lambda$6716.4 		& 2.6(0.2)       & 2.4(0.4)   & 2.6(0.7)   &25(4)	&23(5)       &25(14) \\
$[$\ion{S}{2}$]$$\lambda$6730.8 		& 2.9(0.2)       & 3.2(0.4)   & 3.4(0.7)   &28(5)	&31(6)       &33(14) \\
\enddata 
\tablenotetext{\dag}{~Lines discovered in this work.}
\tablecomments{Errors are given in parenthesis.}
\label{t_flux}
\end{deluxetable} 

%% file: t_masses.tex
\begin{table}
\footnotesize
\begin{center}

\centering \caption{Atomic and ionized mass of M\,2-56 \label{t_masses}
}

\begin{tabular}{ccccccccc}
\hline
\hline
Component    &F$_{H\alpha}$ & F$_{\rm [OI]}$                        & $<$A$_{V}$$>$  & $<$n$_{e}$$>$     & M$_{H^{+}}^{H_{\alpha}}$    & M$_{H}^{\rm [OI]}$ \\      
	     & (10$^{-15}$\,erg\,s$^{-1}$\,cm$^{-2}$) & (10$^{-15}$\,erg\,s$^{-1}$\,cm$^{-2}$) & (mag) &(cm$^{-3}$)  &(M$_{\odot}$) & (M$_{\odot}$)  \\  
\hline
\hline
OLs          &5.7($\pm$0.2) & 2.0($\pm$0.6) &2	 & 200                & 5.3$\times$10$^{-4}$     & 1.5$\times$10$^{-4}$ \\ 
ILs	     &9.3($\pm$0.2) & 5.4($\pm$0.4) &3	 & 1500               & 8.2$\times$10$^{-5}$   & 3.8$\times$10$^{-5}$ \\
F1-wind	     &8.0($\pm$0.2) & -- &$\ga$3 ($\pm$12)&$>$2.6$\times$10$^4$& 4.1$\times$10$^{-6}$\tablenotemark{\dag} ($<$4.2$\times$10$^{-3}$) & -- \\ 
\hline
\hline
\tablenotetext{\dag}{Particularly uncertain. See text in \S\,\ref{mass}}
\end{tabular} 
\end{center}
\end{table}

%% file: ms.bbl
\begin{thebibliography}{}

\bibitem[Aaquist \& Kwok(1990)]{ak90} Aaquist, O.~B., \& Kwok, S.\ 1990, \aaps, 84, 229 

\bibitem[Alcolea et al.(2007)]{alc07} Alcolea, J., Neri, R., \& Bujarrabal, V.\ 2007, \aap, 468, L41 

\bibitem[Balick \& Frank(2002)]{bal02} Balick, B., \& Frank, A.\ 2002, \araa, 40, 439 

\bibitem[Bloecker(1995)]{blo95} Bloecker, T.\ 1995, \aap, 299, 755 

\bibitem[Bujarrabal et al.(2001)]{buj01} Bujarrabal, V., Castro-Carrizo, A., Alcolea, J., \& S{\'a}nchez Contreras, C.\ 2001, \aap, 377, 868 

\bibitem[Cardelli et al.(1989)]{car89} Cardelli, J.~A., 
Clayton, G.~C., \& Mathis, J.~S.\ 1989, \apj, 345, 245 

\bibitem[Castro-Carrizo et al.(2002)]{cc02} Castro-Carrizo, A., Bujarrabal, V., S{\'a}nchez Contreras, C., Alcolea, J., \& Neri, R.\ 2002, \aap, 386, 633 

\bibitem[Cohen 
\& Kuhi(1977)]{coh77} Cohen, M., \& Kuhi, L.~V.\ 1977, \pasp, 89, 829 

\bibitem[Cox et al.(2000)]{cox00} Cox, P., Lucas, R., Huggins, P.~J., Forveille, T., Bachiller, R., Guilloteau, S., Maillard, J.~P., \& Omont, A.\ 2000, \aap, 353, L25 


\bibitem[Goodrich(1991)]{goo91} Goodrich, R.~W.\ 1991, \apj,  376, 654 (G91)

\bibitem[Guerrero et al.(2001)]{gue01} Guerrero, M.~A., 
Miranda, L.~F., Chu, Y.-H., Rodr{\'{\i}}guez, M., 
\& Williams, R.~M.\ 2001, \apj, 563, 883 

\bibitem[Gurzadyan(1997)]{gur97} Gurzadyan, G.~A.\ 1997, 
The Physics and Dynamics of Planetary Nebulae, Springer-Verlag Berlin
Heidelberg New York.~ Also Astronomy and Astrophysics Library.


\bibitem[Hartigan et al.(1987)]{har87} Hartigan, P., Raymond, 
J., \& Hartmann, L.\ 1987, \apj, 316, 323 


\bibitem[Hartigan et al.(1994)]{har94} Hartigan, P., Morse, 
J.~A., \& Raymond, J.\ 1994, \apj, 436, 125 

\bibitem[Imai et al.(2002)]{ima02} Imai, H., Obara, K., 
Diamond, P.~J., Omodaka, T., \& Sasao, T.\ 2002, \nat, 417, 829 

\bibitem[Imai(2007)]{ima07} Imai, H.\ 2007, IAU Symposium, 
242, 279 

\bibitem[van Hoof et al.(1997)]{hoof97} van Hoof, P.~A.~M., 
Oudmaijer, R.~D., \& Waters, L.~B.~F.~M.\ 1997, \mnras, 289, 371 

\bibitem[Howarth(1983)]{how83} Howarth, I.~D.\ 1983, \mnras, 203, 301 

\bibitem[Kwok(2000)]{kwo00} Kwok, S.\ 2000, The origin and 
evolution of planetary nebulae / Sun Kwok.~Cambridge ; New York :
Cambridge University Press, 2000.~(Cambridge astrophysics series ; 33)

\bibitem[Mendoza(1983)]{men83} Mendoza, C.\ 1983, IAU Symp.~103: 
Planetary Nebulae, 103, 143

\bibitem[Miranda et al.(1999)]{mir99} Miranda, L.~F., Guerrero, M.~A., \& Torrelles, J.~M.\ 1999, \aj, 117, 1421 


\bibitem[Miranda et al.(2006)]{mir06} Miranda, L.~F., Ayala, S., V{\'a}zquez, R., 
\& Guill{\'e}n, P.~F.\ 2006, \aap, 456, 591 

\bibitem[Osterbrock \& Ferland(2006)]{ost06} Osterbrock, D.~E., \& Ferland, G.~J.\ 2006, Astrophysics of gaseous nebulae and active galactic nuclei, 2nd.~ed.~by D.E.~Osterbrock and G.J.~Ferland.~Sausalito, CA: University Science Books, 2006,  

\bibitem[Raymond(1979)]{ray79} Raymond, J.~C.\ 1979, \apjs, 39, 1 


\bibitem[Reipurth et al.(2000)]{rei00} Reipurth, B., Yu, K., 
Heathcote, S., Bally, J., \& Rodr{\'i}guez, L.~F.\ 2000, \aj, 120,
1449.

\bibitem[Riera et al.(2006)]{rie06} Riera, A., Binette, L., \& Raga, A.~C.\ 2006, \aap, 455, 203 

\bibitem[Sahai \& Trauger(1998)]{st98} Sahai, R., \& Trauger, J.~T.\ 1998, \aj, 116, 1357 

\bibitem[Sahai et al.(1998)]{sah98} Sahai, R., et al.\ 1998, \apj, 493, 301  

\bibitem[Sahai et al.(2007)]{sahs07} Sahai, R., Morris, M., 
S{\'a}nchez Contreras, C., \& Claussen, M.\ 2007, \aj, 134, 2200 

\bibitem[S{\'a}nchez Contreras et al.(2008)]{san08} S{\'a}nchez Contreras, C., Sahai, R., Gil de Paz, A., 
\& Goodrich, R.\ 2008, \apjs, 179, 166  

\bibitem[S{\'a}nchez Contreras et al.(2007)]{san07} S{\'a}nchez Contreras, C., Le Mignant, D., Sahai, R., Gil de Paz, A., 
\& Morris, M.\ 2007, \apj, 656, 1150  

\bibitem[S{\'a}nchez Contreras et al.(2002)]{san02} 
S{\'a}nchez Contreras, C., Sahai, R., \& Gil de Paz, A.\ 2002, \apj, 578, 269 

\bibitem[S{\'a}nchez Contreras \& Sahai(2001)]{san01} S{\'a}nchez Contreras, C., \& Sahai, R.\ 2001, \apjl, 553, L173 


\bibitem[S{\'a}nchez Contreras et al.(2000)]{san00} S{\'a}nchez Contreras, C., Bujarrabal, V., Miranda, L.~F., \& Fern{\'a}ndez-Figueroa, M.~J.\ 2000, \aap, 355, 1103  


\bibitem[Si{\'o}dmiak et al.(2008)]{sio08} Si{\'o}dmiak, N., 
Meixner, M., Ueta, T., Sugerman, B.~E.~K., Van de Steene, G.~C., 
\& Szczerba, R.\ 2008, \apj, 677, 382   


\bibitem[Solf(2000)]{solf00} Solf, J.\ 2000, \aap, 354, 674   

\bibitem[Su{\'a}rez et al.(2009)]{sua09} Su{\'a}rez, O., G{\'o}mez, J.~F., Miranda, L.~F., Torrelles, J.~M., G{\'o}mez, Y., Anglada, G., \& Morata, O.\ 2009, \aap, 505, 217 


\bibitem[Trammell et al.(1993)]{tra93} Trammell, S.~R., 
Dinerstein, H.~L., \& Goodrich, R.~W.\ 1993, \apj, 402, 249 


\bibitem[Ueta et al.(2007)]{ue07} Ueta, T., Murakawa, K., 
\& Meixner, M.\ 2007, \aj, 133, 1345 


\bibitem[Vel{\'a}zquez et al.(2004)]{vel04} Vel{\'a}zquez, P.~F., Riera, A., \& Raga, A.~C.\ 2004, \aap, 419, 991 

\bibitem[Welch et al.(1999)]{wel99} Welch, C.~A., Frank, A., Pipher, J.~L., Forrest, W.~J., \& Woodward, C.~E.\ 1999, \apjl, 522, L69. 

\end{thebibliography}
